\documentclass[aps,prd,superscriptaddress,nofootinbib,eqsecnum,twocolumn]{revtex4-1}

\pdfoutput=1

\usepackage{graphicx}  % needed for figures
\usepackage{dcolumn}   % needed for some tables
\usepackage{amssymb}   % for math
\usepackage{xcolor}
\usepackage{graphics} 
\usepackage{hyperref}
\usepackage{amsfonts}
\usepackage{latexsym}
\usepackage{slashed}
\usepackage{bm}  % To make greek letters in bold face {\bm \Phi}
\usepackage[normalem]{ulem} %tabela
\usepackage{amsmath,mathtools,mathrsfs,bbm,amssymb} 
\usepackage{bbold}
\usepackage{subfigure}
\usepackage{array}
\usepackage{wrapfig}
\usepackage{multirow}
\usepackage{tabularx}
\usepackage{float}

\begin{document}

\title{Testing the equivalence between the planar Gross-Neveu and Thirring models at $N=1$}

\author{Everlyn Martins} \email{everlyn.martins@posgrad.ufsc.br}
\affiliation{Departamento de F\'{\i}sica, Universidade Federal de Santa
  Catarina, Florian\'{o}polis, SC 88040-900, Brazil}

\author{Y. M. P. Gomes} \email{yurimullergomes@gmail.com}
\affiliation{Departamento de F\'{\i}sica Te\'{o}rica, Universidade do Estado do Rio de Janeiro, Rio de Janeiro, RJ 20550-013, Brazil}
 
\author{Marcus Benghi Pinto} \email{marcus.benghi@ufsc.br}
\affiliation{Departamento de F\'{\i}sica, Universidade Federal de Santa
  Catarina, Florian\'{o}polis, SC 88040-900, Brazil}
  
\author{Rudnei O. Ramos} \email{rudnei@uerj.br}  
\affiliation{Departamento de F\'{\i}sica Te\'{o}rica, Universidade do Estado do Rio de Janeiro, Rio de Janeiro, RJ 20550-013, Brazil}

%%%%%%%%%%%%%%%%%%%%%%%%%%%%%%%%%%%%%%%%%%%%%%%%% 
\begin{abstract} 

It is known that the Fierz identities predict that the Gross-Neveu and
Thirring models should be equivalent when describing systems composed
of a single fermionic flavor, $N=1$. Here, we consider the planar
version of both models within the framework of the optimized
  perturbation theory at the two-loop level, in order
to verify if the predicted equivalence emerges explicitly when
different temperature and density regimes are
considered.  At vanishing
densities, our results indicate that both models indeed describe
exactly the same thermodynamics, provided that $N=1$. However, at
finite chemical potentials we find that the $N=1$ Fierz equivalence no
longer holds. After examining the relevant thermodynamic potentials, we have
identified the contributions which lead to this puzzling
discrepancy. {}Finally, we discuss different frameworks in which this
(so far open)  problem could be further understood and eventually
circumvented. 

\end{abstract}
%%%%%%%%%%%%%%%%%%%%%%%%%%%%%%%%%%%%%%%%%%%%%%%%%

\maketitle 
 
%%%%%%%%%%%%%%%%%%%%%%%%%%%%%%%%%%%%%%%%%%%%%%%%%
\section{Introduction}
 
Relativistic four-fermion theories are widely used to describe
different physical scenarios in condensed matter and particle
physics. In the latter case, the Nambu--Jona-Lasinio (NJL)~\cite{Nambu:1961tp,Nambu:1961fr} model  has been successfully used as an effective model to describe the QCD
chiral phase transition. As far as condensed matter is concerned, the
planar Gross-Neveu (GN)~\cite{Gross:1974jv}  and
Thirring~\cite{Thirring:1958in} models have been employed to describe
low-energy electronic properties of materials like
graphene~\cite{Hands:2008id,Ebert:2015hva,Ebert:2018dzs},
high-temperature
superconductors~\cite{Zhukovsky:2017hzo,Klimenko:2012qi}, Weyl
semimetals~\cite{Gomes:2021nem,Gomes:2022dmf,Gomes:2023vvu}, among
many other
systems~\cite{Caldas:2008zz,Caldas:2009zz,Ramos:2013aia,Klimenko:2013gua,Khunjua:2021fus,Gubaeva:2022feb,Khunjua:2022kxf}. 

The basic difference between the GN and Thirring models arises from
the matrix structure of the four-fermion channel, $(\bar{\psi}_k
\Gamma \psi_k)^2$ ($k=1,...,N$), considered in each case. This channel
has a scalar structure ($\Gamma \equiv\mathbbm{1}$) in the GN case,
while a vector structure ($\Gamma \equiv \gamma^\mu$)  is adopted
within the  Thirring Lagrangian density. Despite this important
physical difference, the use of Fierz
identities~\cite{Fierz:1937wjm,Nieves:2003in} allows us to conclude
that both models turn out to be equivalent when describing systems
composed by a single fermionic flavor, $N=1$ (see,
e.g. Ref.~\cite{Rosenstein:1988zf} and, for a recent discussion,
Ref. \cite{Wipf:2022hqd}). This result offers an additional
opportunity to explicitly test nonperturbative methods which are able
to incorporate finite $N$ corrections.  In this context, it is worth
to mention that some of the traditional  analytical
methods used to study these models, such as the large-$N$ (or
mean-field)  approximation,   cannot reliably accommodate for small
values of $N$, which precludes in explicitly verifying the equivalence
between the models. This represents a special issue when trying to compare the
thermodynamical properties of both models. In addition, one could recur to
lattice simulations in order to explicitly verify the equivalence
predicted by means of the Fierz identities. However, within the domain
of finite fermionic densities, these numerical simulations are plagued
by the well-documented sign
problem~\cite{Karsch:2001cy,Muroya:2003qs}. This fact produces  an
unfortunate situation, since the density is usually crucial for the
description of many condensed matter systems. To circumvent these
difficulties, one must resort to alternative analytical
nonperturbative approximations that are capable of going beyond the
large-$N$ limit. In this context, the  {\it optimized perturbation
  theory} (OPT)~\cite{Okopinska:1987hp,Duncan:1988hw} (see also, e.g.,
Ref.~\cite{Yukalov:2019nhu} for a recent review), offers a convenient
and simple enough framework to study such a problem since, within this 
method, contributions of order $1/N$ typically already appear at its
first nontrivial order. 
 
The OPT has been very successful in the description of phase
transitions within four-fermion theories such as the ones described by
the NJL and GN
models~\cite{Kneur:2007vj,Kneur:2007vm,Kneur:2010yv,Kneur:2013cva}. Here,
this approximation will be employed to describe the thermodynamic
potential of the 2+1 $d$ GN and Thirring models including the crucial first $1/N$ contributions. Although the OPT results for the GN case to be considered
here were originally obtained in
Refs.~\cite{Kneur:2007vj,Kneur:2007vm}, the present work represents
the first OPT application to the Thirring model as far as we are aware of. As we shall
demonstrate, the expected equivalence between the models can be
explicitly confirmed at all temperatures and vanishing densities. On
the other hand, at least at the first order in the OPT method considered here, the equivalence
does not seem to hold exactly when finite chemical potential values
are considered. We offer a discussion about the possible reasons for this
unexpected result. 

The manuscript is organized as follows. In Sec.~\ref{section2}, we present
both models considered in this paper, namely, the GN and Thirring models. In Sec.~\ref{section3}, we obtain the OPT Lagrangian
densities for the two models. The corresponding effective potentials for
each model are presented in Sec.~\ref{section4}. The optimization
procedure is implemented in Sec.~\ref{section5}. The thermodynamical
results are presented and discussed in
Sec.~\ref{section6}. {}Finally, in Sec.~\ref{conclusions} we present
our concluding remarks.  Two appendixes are also included to show some
of the more technical details.
 
%%%%%%%%%%%%%%%%%%%%%%%%%%%%%%%%%%%%%%%%%%%%%%%%%
\section{The Models}
\label{section2}

The GN~\cite{Gross:1974jv} and Thirring~\cite{Thirring:1958in} models can be described using a unified
notation, $\Gamma$, which represents the identity for the GN
model and $\gamma^\mu$ for the Thirring model. The Lagrangian density
for a fermion field $\psi_k \ (k=1, \ldots, N)$ describing both models can then be written as
\begin{equation}
\mathcal{L} = \bar{\psi}_k(i \slashed{\partial}) \psi_k + m_f
\bar{\psi}_k \psi_k + \frac{\lambda}{2N}(\bar{\psi}_k \Gamma \psi_k)^2,
\label{Lgeral}
\end{equation}
where the summation over fermionic species, $\bar{\psi}_k \Gamma
\psi_k = \sum_{k=1}^N \bar{\psi}_k \Gamma \psi_k$, is implied. When
$m_f = 0$, the Lagrangian density given by Eq. (\ref{Lgeral}) is
invariant under the discrete chiral symmetry transformation $\psi
\rightarrow \gamma_5 \psi$. At finite temperatures and
densities, the grand partition function is
\begin{equation}
Z(\beta, \mu) = \operatorname{Tr} \exp [-\beta(H - \mu Q)],
\end{equation}
where $\beta$ is the inverse temperature, $\mu$ the chemical
potential, $H$ the Hamiltonian, and $Q = \int d^2x \; \bar{\psi}_k
\gamma_0 \psi_k$ is the conserved charge. In the Euclidean formalism
and in terms of functional integration
over the fermion fields, we have that
\begin{equation}
Z(\beta, \mu) = \int \prod_{k=1}^N D \bar{\psi}_k D \psi_k \exp
\left\{-S_E\left[\bar{\psi}_k, \psi_k\right]\right\},
\end{equation}
with the Euclidean action given by
\begin{eqnarray}
S_E\left[\bar{\psi}_k, \psi_k\right] &=& \int_0^\beta d \tau \int d^2x
\left[\bar{\psi}_k \left(\slashed{\partial} + \mu \gamma_0 -
  m_f\right) \psi_k  \right.  \nonumber \\ &-& \left. \frac{\lambda}{2
    N}(\bar{\psi}_k \Gamma \psi_k)^2\right],
\end{eqnarray}
and fermion fields satisfying anti-periodic boundary conditions:
$\psi_k(x, \tau) = -\psi_k(x, \tau + \beta)$.

The work in Ref.~\cite{Wipf:2022hqd} has employed the Fierz identities
to show the equivalence between the $N=1$ Thirring and
GN models in both $2d$ and $3d$. The {}Fierz identities are
mathematical transformations that rearrange products of spinor
bilinears into different combinations, revealing underlying symmetries
and equivalences between seemingly distinct interaction
terms~\cite{Fierz:1937wjm,Nieves:2003in}. This equivalence is
expressed here as
\begin{equation}
\frac{\lambda}{2 N}\left(\bar{\psi} \gamma^\mu \psi\right)^2 =
-\frac{d \lambda}{2N} (\bar{\psi} \psi)^2, \quad N=1 ,
\label{equivalence}
\end{equation}
where $d$ represents the spacetime dimension, which can be either
$d=2$ or $d=3$. This result highlights that, for a single fermion
flavor, the interaction term in the Thirring model, which involves a
vector current-current interaction, is mathematically equivalent to
the scalar-scalar interaction term in the GN model. Such duality arises due to the specific properties of the spinor
components in both two and three dimensions. Since our work focuses on
the three-dimensional case with $N=1$, this equivalence is directly
relevant to our study.

%%%%%%%%%%%%%%%%%%%%%%%%%%%%%%%%%%%%%%%%%%%%%%%%%
\section{Interpolated Theories}
\label{section3}

The implementation of the OPT technique follows a deformation of the
original Lagrangian density by adding and subtracting a Gaussian term
$(1-\delta)\eta \overline{\psi}_k \psi_k$, while the coupling constant,
$\lambda$, which parametrizes the four-fermion vertex becomes $\delta
\lambda$.  A given physical quantity can then be evaluated as a
perturbative series in powers of $\delta$. Once the series is
truncated at a given order-$\delta^k$ the bookkeeping parameter is set
to its original value, $\delta=1$, while the mass parameter ($\eta$)
is usually optimized in a variational fashion. In general, this
procedure allows for a resummation of the original perturbative
series, generating nonperturbative
results~\cite{Pinto:1999py,Pinto:1999pg,Farias:2008fs,Rosa:2016czs,Farias:2021ult}. One
of the advantages of such a method is that originally massless
theories, such as the ones to be considered here, are well-behaved in
the infrared limit since the variational mass parameter ($\eta$)
naturally regularizes all integrals at vanishing momenta. Another
welcome feature is that the actual evaluations are carried out using
the usual perturbation theory framework (including the renormalization
procedure).

The application of the OPT method to the Thirring model follows a
similar procedure to the one used for the GN model considered in many previous
works~\cite{Kneur:2007vj,Kneur:2007vm,Kneur:2013cva}. The modified
Lagrangian density for this model, including a linear
interpolation with the fictitious parameter $\delta$ of the OPT
prescription, is given by
\begin{equation}
\mathcal{L}_\text{Th} = \bar{\psi}_k(i \slashed{\partial}) \psi_k +
(1-\delta) \eta \bar{\psi}_k \psi_k + \frac{\delta \lambda}{2 N}
\left(\bar{\psi}_k \gamma^\mu \psi_k \right)^2.
\label{LTh}
\end{equation}

As $\delta$ varies from 0 to 1, the theory interpolates from free
fermions to the original interacting theory. This formulation allows
the application of the OPT method to derive nonperturbative results
for the Thirring model, similar to the approach taken for the GN
model \cite{Kneur:2007vj,Kneur:2007vm,Kneur:2013cva}. The Lagrangian density (\ref{LTh}) can also more conveniently
be expressed in terms of an auxiliary vector field $v^\mu$ by adding the
term
\begin{equation}
-\frac{\delta N}{2 \lambda}\left(v^\mu + \frac{\lambda}{N}
\bar{\psi}_k \gamma^\mu \psi_k\right)^2\,,
\end{equation}
to it. This leads us to the interpolated model,
\begin{equation}
\mathcal{L}_\text{Th} = \bar{\psi}_k (i \slashed{\partial}) \psi_k -
\bar{\psi}_k \left[\eta -  \delta \left(\eta - \slashed{v}
  \right)\right] \psi_k - \frac{\delta N}{2\lambda} v^\mu v_\mu,
\label{IntThLagrangian}
\end{equation}
where  $\slashed{v} = \gamma^\mu v_\mu$. The original GN 
Lagrangian density, when deformed in the same way, leads
to~\cite{Kneur:2007vm}
\begin{equation}
\mathcal{L}_\text{GN} = \bar{\psi}_k(i \slashed{\partial}) \psi_k -
\bar{\psi}_k [\eta - \delta (\eta - \sigma) ] \psi_k - \frac{\delta
  N}{2 \lambda} \sigma^2,
\label{IntGNlagrag}
\end{equation}
where $\sigma$ is an (auxiliary) scalar field.  Note that the
Euler-Lagrange equations for Eqs.~(\ref{IntThLagrangian}) and
(\ref{IntGNlagrag}) set the auxiliary fields to $v^\mu = -
({\lambda}/{N}) \bar{\psi}_k \gamma^\mu \psi_k$ and $\sigma = -
({\lambda}/{N}) \bar{\psi}_k  \psi_k$, respectively.

{}From the interpolated Lagrangian densities,
Eqs.~(\ref{IntThLagrangian}) and (\ref{IntGNlagrag}), we can derive
the new {}Feynman rules in Minkowski space: each Yukawa vertex carries a
factor  $-i \delta \Gamma$, while both auxiliary field propagators
read $-i \lambda /(N \delta)$. At the same time, the fermion propagator now reads $i (\slashed {p} - \eta + i\epsilon)^{-1}$ and a novel (quadratic) vertex contributes with $-i \delta \eta$.

The optimization process starts by
using some variational principle, such as the widely used 
principle of minimal sensitivity (PMS)~\cite{Stevenson:1981vj}. The
PMS criterion consists of applying the following variational condition
to the physical quantity under consideration, which in the present
case is taken to be the effective potential. This means that the optimal variational mass must satisfy the relation
\begin{equation}
\left.\frac{d V_{\rm eff}}{d \eta}\right|_{\overline{\eta}}=0\,,
\label{PMSopt}
\end{equation}
as we shall demonstrate in the next section.

%%%%%%%%%%%%%%%%%%%%%%%%%%%%%%%%%%%%%%%%%%%%%%%%%%%%%%%%%%%%%%%%%%%%%%%%%%%%%%
\section{The effective potentials in the OPT scheme}
\label{section4}

%%%%%%%%%%%%%%%%%%%%%%%%%%%%%%%%%%%%%%%%%%%%%%%%%
\begin{figure}[!htb]
\centering \includegraphics[width=8cm]{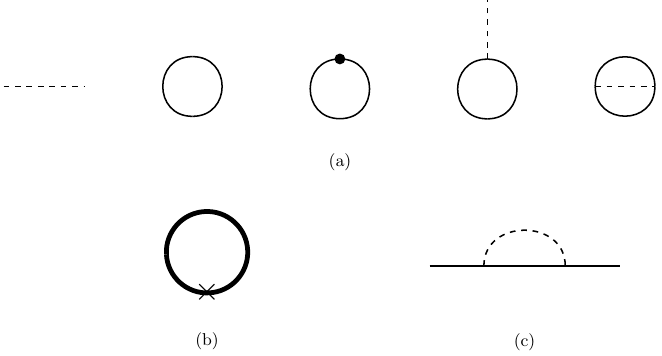}
\caption{(a) Diagrams contributing to the effective potential at
  order-$\delta$. The external dashed line represents the auxiliary
  fields $\sigma$ (GN), or $v^\mu$ (Thirring). The internal dashed
  line represents the respective propagators associated with each
  auxiliary field. The thin solid lines represent the (bare) fermion
  propagator in terms of $\eta$. The Yukawa vertex is proportional to
  $\gamma_\mu$ (Thirring) or to the identity matrix $\mathbbm{1}$
  (GN). (b) Diagram representing the quark condensate and the fermion number density when ``x" respectively represents the identity or
  $\gamma_0$. The thick continuous line represents the dressed (optimized) fermion propagator, in terms of $\overline \eta$ (c) The exchange (self-energy) fermionic
  one-loop contribution of order $\delta/N$.}
\label{Fig1}
\end{figure}
%%%%%%%%%%%%%%%%%%%%%%%%%%%%%%%%%%%%%%%%%%%%%%%%%

Let us start by recalling that the {}Fierz equivalence relation,
Eq.~(\ref{equivalence}), requires that the four-fermion vertices
describing the Thirring and GN models have opposite signs. Since we are
interested in studying the breaking/restoration of chiral symmetry, it is convenient to perform the replacement $\lambda \to
|\lambda|$ in Eq.~(\ref {IntThLagrangian}) and  $\lambda \to
-|\lambda|$ in Eq.~(\ref{IntGNlagrag}). These particular choices can
be justified by recalling that, within the planar GN model, chiral
symmetry breaking only occurs when the coupling is
negative\footnote{An exception to this rule occurs in the presence of
a magnetic field, as detailed in Ref.~\cite{Kneur:2013cva} and the
references therein.}. {}Furthermore, since the coupling $\lambda$ has
canonical dimension [-1] as well as to be consistent with the convention
adopted in previous applications, we introduce the scale $\Lambda
\equiv \pi / |\lambda|$.

The terms contributing to the effective potential for the Thirring
model, at first order in the OPT approximation, are illustrated in
{}Fig.~\ref{Fig1}(a). Then, in terms of the
scale $\Lambda$, one explicitly obtains
\begin{eqnarray}
\lefteqn{\frac{V_{\rm eff}^\text{Th}\left(v_\mu, \eta\right)}{N} =
  \delta\frac{\pi}{2 \Lambda} v^\mu v_\mu  } \nonumber \\ &&+ i \int
\frac{d^d p}{(2 \pi)^d}  \operatorname{Tr} \ln
\left(\slashed{p}-\eta\right) +  i  \int \frac{d^d p}{(2 \pi)^d}
\operatorname{Tr} \left[\frac{\delta \eta }{\slashed{p} -  \eta+i
    \epsilon}\right]   \nonumber \\ &&- i  \int \frac{d^d p}{(2
  \pi)^d}  \operatorname{Tr} \left[\frac{\delta \slashed{v}
  }{\slashed{p} -  \eta+i \epsilon}\right]  \nonumber \\ &&+ \delta
\frac{ \pi}{2N \Lambda} \int \frac{d^d p}{(2 \pi)^{d}} \frac{d^d q}{(2
  \pi)^{d}}
\operatorname{Tr}\left[\frac{i}{\slashed{p}-\eta+i\epsilon} \gamma_\mu
  \frac{i}{\slashed{q} -\eta+i \epsilon} \gamma^\mu\right].  \nonumber
\\
\label{PotEffTh}
\end{eqnarray}

Taking the traces (see  Appendix~\ref{AppendixA}) over the $\gamma_\mu$
matrices in Eq.~(\ref{PotEffTh}) and noting that, due to spatial
translation invariance, only the zeroth component of $v^\mu$ within the
integral contributes to the fourth term, one can  finally rewrite
Eq.~(\ref {PotEffTh}) as 
\begin{eqnarray}
\lefteqn{ \frac{V_{\rm eff}^{\rm Th}\left(v_\mu, \eta\right)}{N}
  =\delta\frac{\Lambda}{2 \pi} v^\mu v_\mu +  2 i\int \frac{d^d p}{(2
    \pi)^d} \ln (p^2-\eta^2) } \nonumber \\ &&+ 4 \delta \eta^2 i \int
\frac{d^d p}{(2 \pi)^d} \frac{1}{p^2-\eta^2+i \epsilon}  \nonumber
\\ &&- 4 \delta v_0 i \int \frac{d^d p}{(2 \pi)^d}
\frac{p_0}{p^2-\eta^2+i \epsilon}  \nonumber\\ & & + 6
\delta\frac{\pi}{\Lambda N}  \eta^2\left[i \int \frac{d^d
    p}{(2 \pi)^d} \frac{1}{p^2-\eta^2+i \epsilon}\right]^2  \nonumber
\\ &&- 2 \delta\frac{\pi}{\Lambda N} \left[i \int \frac{d^d
    p}{(2 \pi)^d} \frac{p_0}{p^2-\eta^2+i \epsilon}\right]^2.
\label{Veff}
\end{eqnarray}
Within the imaginary time formalism, adopted here, the control
parameters represented by the temperature ($T$) and chemical potential
($\mu$) are introduced by defining the relativistic momentum to be
$p=\left(p_0=i\left(\omega_n-i \mu\right), \mathbf{p}\right)$ where
$\omega_n=(2 n+1) \pi T$ ($n=0, \pm 1, \pm 2, \ldots$) represents the
Matsubara frequencies for fermions. The momentum integrals in
Eq.~(\ref{Veff} can be expressed like in Eq.~(\ref{momint}).  Carrying
out the summation over the Matsubara's frequencies (see
Appendix~\ref{AppendixB} for details), one finally gets the more
compact result
\begin{eqnarray}
\frac{V_{\rm eff}^{\rm Th}\left(v_\mu, \eta \right)}{N} &=&
\delta\frac{\Lambda}{2 \pi} v^\mu v_\mu + 2 \mathcal{X}_p (\eta) + 4
\delta \eta^2 \mathcal{Y}_p (\eta)  \nonumber \\ &-&4 \delta v_0
\mathcal{W}_p (\eta)  \nonumber \\ &+& 2 \delta\frac{\pi}{\Lambda
  N}\left[ 3\eta^2 \mathcal{Y}_p^2 (\eta) - \mathcal{W}_p^2
  (\eta)\right],
\label{Th3dPot}
\end{eqnarray}
where the thermal integrals $\mathcal{X}_p, \; \mathcal{Y}_p$ and
$\mathcal{W}_p$  are, respectively, given
by Eqs.~(\ref{XTmu}), (\ref{YTmu}) and (\ref{WTmu}) explicitly  presented in  Appendix~\ref{AppendixB}. 

{}Following a similar procedure, the  GN effective potential can be
written as~\cite{Kneur:2007vm} 
\begin{eqnarray}
\frac{V_{\rm eff}^{\rm GN}\left(\sigma, \eta\right)}{N} &=&
-\delta\frac{\Lambda}{2 \pi} \sigma^2+ 2 \mathcal{X}_p (\eta) +
4\delta \eta \left(\eta-\sigma\right) \mathcal{Y}_p (\eta)  \nonumber
\\ &-&2 \delta\frac{\pi}{\Lambda N}\eta^2  \mathcal{Y}_p^2 (\eta) -  2
\delta\frac{\pi}{\Lambda N}\mathcal{W}_p^2 (\eta).
\label{GN3dPot}
\end{eqnarray}

The optimized chiral condensate and the particle number density, valid
for both models,  are represented by the diagram of {}Fig.~\ref{Fig1}(b). Performing an explicit evaluation, one obtains
\begin{equation}
\frac{\langle \bar{\psi}_k \psi_k \rangle(\mu,T)}{N} = - 4
\overline{\eta} \mathcal{Y}_p(\overline{\eta}) \;,
\label{ChiralCondDef}
\end{equation}
and
\begin{equation}
\frac{\langle \psi^{\dagger}_k \psi_k \rangle(\mu,T)}{N}  =  - 4
\mathcal{W}_p(\overline{\eta}) \; .
\label{DensDef}
\end{equation}
Notice that both quantities carry nonperturbative information through
$\overline{\eta}$, which by satisfying the PMS criterion,
Eq.~(\ref{PMSopt}), becomes a function of the coupling. The
mathematical structure of these two  quantities allows us to
understand  the  physical origin of the contributions entering the
effective potentials defined by Eqs.~(\ref {Th3dPot}) and (\ref
{GN3dPot}). Let us do this analysis starting with the term proportional to
$\mathcal{X}_p (\eta)$, which in   both cases still contributes at
large $N$. Mathematically, as one can easily check, $ d \mathcal{X}_p
(\eta)/d \eta \sim  \mathcal{Y}_p$. Physically, this term represents a
contribution which is similar to the one generally found in a free
fermionic system composed by (quasi) particles of mass
$\eta$. Regarding finite $N$ effects, one may notice a major difference
in the structure of both effective potentials since in the GN case the
contributions proportional to $\mathcal{W}_p$ (hence, to the number
density) are $1/N$ suppressed, contrary to what happens in the
Thirring case. Then, as $N\to 1$ the two structures become more
similar as one can observe by comparing the thermodynamic
potentials. The effective and thermodynamic potentials are related by
$\Omega_\text{Th}(\mu, T) = V_{\rm eff}^\text{Th}(\mu, T,
\overline{\eta}_\text{Th}, \overline{v}_0)$ and $\Omega_\text{GN}(\mu,
T) = V_{\rm eff}^\text{GN}(\mu, T, \overline{\eta}_\text{GN},
\overline{\sigma})$. As already discussed, while $\overline{\eta}$ can
be determined by the PMS criterion, the auxiliary fields vacuum
expectation values are fixed by the gap equations. In the GN case, the
 equation that determines the vacuum expectation value $\overline
\sigma$ is
\begin{equation}
    \frac{dV_{\rm eff}^\text{GN} }{d \sigma} = \frac{\partial V_{\rm
        eff}^\text{GN} }{\partial  \sigma} + \frac{ \partial
      {\overline \eta} }{\partial  \sigma}\frac{\partial V_{\rm
        eff}^\text{GN} }{\partial  {\overline \eta}} \equiv 0  .
        \label{eqdVeff}
\end{equation}
Due to the PMS condition,  the last derivative on the right-hand side
of Eq.~(\ref{eqdVeff}) vanishes. Hence,  the gap equation yields 
\begin{equation}
\overline{\sigma} = - 4 \frac{\pi}{\Lambda} \overline{\eta}_\text{GN}
\mathcal{Y}_p(\overline{\eta}_\text{GN} )  \equiv \frac{\pi}{\Lambda
  N} \langle \bar{\psi}_k \psi_k \rangle_\text{GN} \,,
\label{SigBarGN}
\end{equation}
which is in agreement with the fact that $\sigma = \pi/(\Lambda N)
\bar{\psi}_k \psi_k$ (see discussion after Eq.~(\ref{IntGNlagrag})).
Applying a similar procedure to the Thirring case, one obtains 
\begin{equation}
\overline{v}_0 = 4 \frac{\pi}{\Lambda} \mathcal{W}_p
(\overline{\eta}_\text{Th}) \equiv -  \frac{\pi}{\Lambda N} \langle
\psi_k^{\dagger} \psi_k \rangle_\text{Th} ,
\label{v0ThGap}
\end{equation}
and $\overline{v}_1 = \overline{v}_2 \equiv 0 $. Note that these
results are in line with the fact that $v^\mu = \pi/(\Lambda
N)\bar{\psi}_k \gamma^\mu \psi_k$, as implied by applying the
Euler-Lagrange equations for $v^\mu$ to Eq.~(\ref
{IntThLagrangian}). Having  obtained the vacuum expectation values (vev) $\overline{\sigma}$ and
$\overline{v}_0$, one can write the thermodynamic potentials in terms
of a generic $\eta$, yet to be optimized, obtaining
\begin{eqnarray}
    \Omega_\text{Th}(\mu, T) &=& 2 \mathcal{X}_p (\eta) -
    \frac{8\pi}{\Lambda}\mathcal{W}_p^2 (\eta) +  4 \eta^2
    \mathcal{Y}_p (\eta)  \nonumber \\  &+& 2 \frac{\pi}{\Lambda N}
    \left [ 3 \eta^2 \mathcal{Y}_p^2 (\eta) - \mathcal{W}_p^2 (\eta)
      \right ]\,,
\end{eqnarray}
and 
\begin{eqnarray}
    \Omega_\text{GN}(\mu, T) &=& 2 \mathcal{X}_p (\eta) +
    \frac{8\pi}{\Lambda}\eta^2 \mathcal{Y}_p^2 (\eta)  + 4 \eta^2
    \mathcal{Y}_p (\eta) \nonumber \\  &-& 2 \frac{\pi}{\Lambda N}
    \left [  \eta^2 \mathcal{Y}_p^2 (\eta) + \mathcal{W}_p^2 (\eta)
      \right ] \,.
\end{eqnarray}
At $N=1$ these equations simplify to
\begin{eqnarray}
    \Omega_\text{Th}(\mu, T) &=& 2 \mathcal{X}_p (\eta)  + 4 \eta^2
    \mathcal{Y}_p (\eta)+\frac{6\pi}{\Lambda}\mathcal{Y}_p^2 (\eta)
    \nonumber \\ &-& \frac{10\pi}{\Lambda} \mathcal{W}_p^2 (\eta)  \,,
\end{eqnarray}
and 
\begin{eqnarray}
    \Omega_\text{GN}(\mu, T) &=& 2 \mathcal{X}_p (\eta) + 4 \eta^2
    \mathcal{Y}_p (\eta)+\frac{6\pi}{\Lambda}\mathcal{Y}_p^2 (\eta)
    \nonumber \\ &-& \frac{2\pi}{\Lambda}\mathcal{W}_p^2 (\eta)  \,.
\end{eqnarray}
Since $\mathcal{W}_p(\eta)=0$ at vanishing densities,  one can immediately
see that, at least in this regime, the models are completely
equivalent as the {}Fierz identities predict. The origin of the
difference appearing at finite densities can be easily tracked down by
noticing that the vector field-dependent contributions to the
effective potential give exactly the extra term which spoils the
expected equivalence. More precisely, 
\begin{equation}
    \frac{{\Lambda \overline{v}_0}^2 }{2 \pi} - 4 {\overline{v}_0}
    \mathcal{W}_p(\eta) = - \frac{8\pi}{\Lambda} \mathcal{W}_p^2(\eta) \,.
\end{equation}
In principle, this apparent structural difference could be compensated, as a consequence of the optimization process, by means of some particular $\overline{\eta}_{\rm GN} \ne \overline{\eta}_{\rm Th}$. This possibility will be explicitly verified in the sequel (Sec. \ref {section6}), just after determining the PMS equations in the next section. 

%%%%%%%%%%%%%%%%%%%%%%%%%%%%%%%%%%%%%%%%%%%%%%%%%%%%%%%%%%%%%%%%%%%%%%%%%%%%%%
\section{Optimization procedure}
\label{section5}

{}For the Thirring model, the PMS condition, Eq.~(\ref{PMSopt}), gives
\begin{equation}
\begin{aligned}
\left.\left\{\left[\eta + \frac{3\pi}{\Lambda N} \eta
  \mathcal{Y}_p(\eta)\right] \left(\mathcal{Y}_p(\eta) + \eta \frac{d
  \mathcal{Y}_p(\eta)}{d \eta}\right)  \right. \right. &
\\ \left. \left. - \left(v_0 + \frac{\pi}{\Lambda
  N}\mathcal{W}_p(\eta)\right) \frac{d \mathcal{W}_p(\eta)}{d
  \eta}\right\}\right|_{\eta=\overline{\eta}_\text{Th}} &= 0 \,.
\end{aligned}
\label{PMSTh}
\end{equation}
As observed in
Refs.~\cite{Kneur:2006ht,Kneur:2007vj,Kneur:2007vm,Kneur:2013cva},
the above optimization relation becomes particularly interesting at
vanishing densities, when the last term in Eq.~(\ref{PMSTh}) does not contribute. In this
case, one obtains 
\begin{equation}
\left.\left(\eta+\frac{3\pi}{\Lambda} \eta
\mathcal{Y}_p(\eta)\right)\left(\mathcal{Y}_p(\eta)+ \eta \frac{d
  \mathcal{Y}_p(\eta)}{d \eta}\right)
\right|_{\eta=\overline{\eta}_\text{Th}}=0 \,,
\label{PMSThmu0}
\end{equation}
where, usually, the solution given by the second term on the left-hand
side in the above equation is discarded on the grounds that it is coupling independent (and
hence, model independent)~\cite{Gandhi:1990yj}. The (physical)
solution given by the first term on the left-hand side of Eq.~(\ref
{PMSThmu0}) allows us to conclude that $\overline{\eta}_\text{Th}=0$
when $N  \to \infty$. Therefore, in this limit, the quark condensate
given by Eq.~(\ref{ChiralCondDef}) also vanishes, preventing the
dynamical breaking of the underlying chiral symmetry observed at the
classical level. On the other hand, any finite value of fermionic
species leads to $\langle \bar{\psi} \psi \rangle \ne 0$, allowing us
to conclude that, at least at first order, the OPT predicts the
critical number of flavors to be $N = \infty$, in accordance with the
Gaussian approximation  result~\cite{Hyun:1994fb}. {}From the
physical point of view, it is often useful to identify the optimal
variational mass with the $\delta/N$ contribution to the self-energy
(represented by the diagram shown in {}Fig.~\ref{Fig1}(c)). This  can be
achieved by remarking that, within the OPT, the explicit evaluation of
this exchange (Fock) type of contribution yields
\begin{equation}
\Sigma_\text{Th}^{\text{exc}}(\overline{\eta}_\text{Th}, T, \mu) =
-\delta \frac{3\pi}{\Lambda N}\overline{\eta}_\text{Th}
\mathcal{Y}_p(\overline{\eta}_\text{Th}) \,,
\label{SelfEnergTh}
\end{equation}
such that, after setting $\delta=1$ in Eq.~(\ref{SelfEnergTh}), one
can identify
\begin{equation}
\overline{\eta}_\text{Th}=\Sigma_\text{Th}^{\text{exc}}(\overline{\eta}_\text{Th},
T, \mu=0) \,.
\label{EtaThmuzero}
\end{equation}
 
{}For the GN model, the PMS relation Eq.~(\ref{PMSopt})  factorizes as
\begin{equation}
\begin{aligned}
\left.\left\{\left[\eta - \sigma - \frac{\pi}{\Lambda N} \eta
  \mathcal{Y}_p(\eta)\right] \left(\mathcal{Y}_p(\eta) + \eta \frac{d
  \mathcal{Y}_p(\eta)}{d \eta}\right)  \right. \right. &
\\ \left. \left. - \frac{\pi}{\Lambda N}  \mathcal{W}_p(\eta) \frac{d
  \mathcal{W}_p(\eta)}{d
  \eta}\right\}\right|_{\eta=\overline{\eta}_\text{GN}} &= 0\,.
\end{aligned}
\label{PMSGN}
\end{equation}
At vanishing densities, Eq.~(\ref{PMSGN}) allows us to illustrate
how the OPT-PMS resummation incorporates finite $N$ contributions in
a nonperturbative fashion. At $\delta=1$, the self-energy exchange
(Fock) term in the GN case can be written as
\begin{equation}
    \Sigma_\text{GN}^{\text{exc}} (\overline{\eta}_\text{GN} , T, \mu)
    = \frac{\pi}{\Lambda N} \overline{\eta}_\text{GN}
    \mathcal{Y}_p(\overline{\eta}_\text{GN} ) \,.
    \label{selfenergyGN}
\end{equation}
Then, when $\mu = 0$, the PMS equation simplifies to
\begin{equation}
\begin{aligned}
\Big[\overline{\eta}_\text{GN} - \sigma -
  \Sigma_\text{GN}^{\text{exc}}(\overline{\eta}_\text{GN}, T,
  \mu=0)\Big]  &
\\ \times\left(\mathcal{Y}_p(\overline{\eta}_\text{GN}) +
\overline{\eta}_\text{GN} \frac{d
  \mathcal{Y}_p(\overline{\eta}_\text{GN})}{d
  \overline{\eta}_\text{GN}}\right) &= 0\,,
\end{aligned}
\label{pmseqGN}
\end{equation}
leading to the simple self-consistent relation
\begin{equation}
\overline{\eta}_\text{GN}=\sigma+\Sigma_\text{GN}^{\text{exc}}(\overline{\eta}_\text{GN}
, T, \mu=0)\,,
\label{consteqGN}
\end{equation}
after discarding the nonphysical solution given by the second term on
the left-hand side in Eq.~(\ref{pmseqGN}).  Then, by combining
Eqs.~(\ref{consteqGN}) and (\ref{SigBarGN}), one also obtains the  relation
\begin{equation}
\overline{\eta}_\text{GN} = \left(1-\frac{1}{4 N} \right)
\overline{\sigma}\,,
\label{EtaBarGNsimple}
\end{equation}
which can be useful when describing the $\mu=0$ scenario.  In the next
section, we  present the solutions for each  optimization equation and
analyze the thermodynamic behavior for both models at different regimes
of temperatures and chemical potentials at different values of $N$. This will
allow us to check how each model  approaches the duality relation when
$N=1$.

%%%%%%%%%%%%%%%%%%%%%%%%%%%%%%%%%%%%%%%%%%%%%%%%%
\section{Results}
\label{section6}

Having obtained the vacuum expectation values of the auxiliary fields,
$\overline \sigma$ and ${\overline v}_0$, as well as the equations for
the optimal variational masses, one can further write down the two
relevant thermodynamic potentials to explicitly analyze the difference
$\Delta \Omega = (\Omega_\text{Th} - \Omega_\text{GN}) $ in three
extreme boundaries of the $T-\mu$ plane for different values of $N$. 

%%%%%%%%%%%%%%%%%%%%%%%%%%%%%%%%%%%%%%%%%%%%%%%%%
\subsection{The $T=0$ and $\mu=0$ case}
\label{SectionMuTZero}

Let us start by considering the case of vanishing temperatures and
densities (vacuum). Using Eq.~(\ref{EtaThmuzero}), together with the
definitions presented in Appendix~\ref{AppendixB}, one obtains
\begin{equation}
    \overline{\eta}_{\text{Th}} = \frac{4 N \Lambda }{3} \, .
    \label{EtaBarMuTZeroTh}
\end{equation}
Then, since $\overline {v}_0=0$ (see Eq.~(\ref{v0ThGap})), the
Thirring thermodynamic potential can be readily written as 
\begin{equation}
    \Omega_\text{Th}(\mu=0, T=0) = -\frac{32 N^3 \Lambda^3}{81 \pi}
    \,.
    \label{FEnergThTMuZeroBar}
\end{equation}
Regarding the GN case, the results obtained in the previous section
lead to 
\begin{equation}
    \overline{\eta}_{\text{GN}} = \frac{\Lambda}{1-\frac{1}{4 N}} \,,
    \label{EtaBarMuTZeroGN}
\end{equation}
which, when combined with Eq.~(\ref{EtaBarGNsimple}), yields the
following thermodynamic potential for the GN model,
\begin{equation}
    \Omega_\text{GN}(\mu=0, T=0) = -\frac{\Lambda^3}{6 \pi
      \left(1-\frac{1}{4 N}\right)^3}\,.
    \label{FEnergGNTMuZeroBar}
\end{equation}
Then, the thermodynamic potential difference between the Thirring and
GN models becomes
\begin{equation}
    \frac{\Delta \Omega (\mu=0, T=0)}{N} =  \frac{N^3 \Lambda^3}{27
      \pi}\left[\left(\frac{3}{4 N-1}\right)^3-1\right] \,,
    \label{DifT0mu0}
\end{equation}
demonstrating that only when $N=1$ one obtains $\Delta \Omega (\mu=0,
T=0) = 0$. This result explicitly confirms that, at least when no
control parameters are present, the OPT procedure at first order is
already sufficient to reproduce the predicted equivalence between the
two different models.

The optimized chiral condensate for the Thirring model can be obtained
from Eq.~\eqref{ChiralCondDef} simply by  replacing $\eta \to
\overline{\eta}_\text{Th} $,
\begin{equation}
\frac{\langle \bar{\psi}_k \psi_k \rangle_\text{Th}}{N} =
\frac{\overline{\eta}_\text{Th}^2}{\pi} = \frac{16 N^2 \Lambda^2}{9}
\;.
\end{equation}
In the same way, the  GN fermion condensate can be written as
\begin{equation}
\frac{\langle \bar{\psi}_k \psi_k \rangle_\text{GN}}{N} =
\frac{\overline{\eta}_\text{GN}^2}{\pi}=  \frac{\Lambda^2}{\pi
  \left(1-\frac{1}{4 N}\right)^2} \;.
\end{equation}
{}For a general number of flavors, the difference between the two
chiral condensates, $\Delta \langle \bar{\psi}_k \psi_k \rangle =
\langle \bar{\psi}_k \psi_k \rangle_\text{Th} - \langle \bar{\psi}_k
\psi_k \rangle_\text{GN}$, is given by
\begin{equation}
\frac{\Delta\langle \bar{\psi}_k \psi_k \rangle}{N} =
\frac{1}{\pi}\left(\frac{4 N
  \Lambda}{3}\right)^2\left[1-\left(\frac{3}{4 N-1}\right)^2\right]
\;,
\end{equation}
{}which, at $N=1$,  simplifies to $\Delta \langle \bar{\psi} \psi
\rangle = 0$.

%%%%%%%%%%%%%%%%%%%%%%%%%%%%%%%%%%%%%%%%%%%%%%%%%
\subsection{The $T\ne 0$ and $\mu = 0$ case}

Let us now consider the case of finite temperatures and vanishing densities. In this regime, the PMS condition,
Eq.~\eqref{EtaThmuzero},  requires  $\overline{\eta}_\text{Th}$
to satisfy
\begin{equation}
\left.\left\{ \frac{4 N \Lambda}{3} -  \left[ |\eta| + 2T \ln \left(1
  + e^{-|\eta| / T}\right) \right] \right\}\right|_{\eta =
  \overline{\eta}_{\text{Th}}} = 0 \, ,
\end{equation}
whose analytical solution is
\begin{equation}
\overline{\eta}_\text{Th}(T)=T \ln \left[\frac{e^{\frac{2 \Lambda
        N}{3 T}}}{2}   \left(\sqrt{e^{\frac{4 \Lambda  N}{3
        T}}-4}+e^{\frac{2 \Lambda  N}{3 T}}\right)-1\right].
\label{EtaBarVsTTh}
\end{equation}
The above equation allows us to evaluate the critical
temperature at which $\overline{\eta}_\text{Th}(T)$ vanishes and that, hence,
signals the point for the chiral symmetry restoration. The result is
\begin{equation}
T_\text{c}^\text{(Th)} = \frac{4 N}{3} \frac{\Lambda}{2 \ln 2} \,.
\label{TcTh}
\end{equation}
Regarding the thermodynamic potential, we note that in this ($\mu=0)$
limit the integral $\mathcal{W}_p(\eta)$ vanishes once again, such
that the gap equation for the vector auxiliary field  simplifies to
$\overline{v}_0 = 0$. It is also clear that using
Eq.~(\ref{EtaBarVsTTh}) to write an explicit equation for the
corresponding thermodynamic potential does not produce any
illuminating analytical result. Nevertheless, as we shall see,
considering the implicit form 
\begin{equation}
\begin{aligned}
\frac{\Omega_\text{Th}(\mu=0, T)}{N} = & \, 2 \mathcal{X}_p
(\overline{\eta}_\text{Th})  + 4 \overline{\eta}_\text{Th}^2
\mathcal{Y}_p (\overline{\eta}_\text{Th}) \\ & + \frac{6 \pi}{N
  \Lambda} \overline{\eta}_\text{Th}^2 \mathcal{Y}_p^2
(\overline{\eta}_\text{Th}),
\end{aligned}
\label{FEnergThMuZeroBar}
\end{equation}
will prove to be sufficient for our comparison purposes.  Now, turning
our attention to the GN case, we notice that at $\mu=0$ the
optimization criterion requires 
\begin{equation}
\left.\left\{ \frac{\Lambda}{1-\frac{1}{4N}} - \left[ |\eta| + 2T \ln
  \left(1 + e^{-|\eta| / T}\right) \right] \right\}\right|_{\eta =
  \overline{\eta}_{\text{GN}}} = 0 \,,
\end{equation} 
from which one obtains
\begin{equation}
\begin{aligned}
\overline{\eta}_\text{GN}(T) &=  \, T \ln \Bigg[\left(\frac{e^{\frac{2
        N \Lambda}{(4N-1) T}}}{2}\right) \left(\sqrt{e^{\frac{4 N
        \Lambda}{(4N-1) T}}-4} \right.  \\ &+ \left.  e^{\frac{2
      \Lambda}{(4N-1) T}}\right) - 1 \Bigg] \, .
\end{aligned}
\label{EtaBarVsTGN}
\end{equation}

One can now determine the corresponding $T_\text{c}$ by setting
$\overline{\eta}_\text{GN}(T)=0$ in Eq.~(\ref{EtaBarVsTGN}). This
yields
\begin{equation}
T_\text{c}^\text{(GN)} = \frac{1}{1-\frac{1}{4N}} \frac{\Lambda}{2 \ln
  2},
\label{TcGN}
\end{equation}
in agreement with Ref.~\cite{Kneur:2007vm}. Notice also that the
well-known large-$N$ result~\cite{Rosenstein:1990nm} is automatically
reproduced upon taking the limit $N \to \infty$ in Eq.~(\ref {TcGN}).
The thermodynamic potential, as a function of
$\overline{\eta}_\text{GN}(T)$, can be written by considering
Eq.~(\ref{EtaBarGNsimple}) for $\overline \sigma$. One then obtains 
\begin{eqnarray}\nonumber
    &&\frac{\Omega_\text{GN}(\mu=0, T)}{N} = 2 \mathcal{X}_p
  (\overline{\eta}_\text{GN}) + 4 \overline{\eta}_\text{GN}^2
  \mathcal{Y}_p (\overline{\eta}_\text{GN}) 
  \\ &&\hspace{+1.cm}+\left(1 - \frac{1}{4N}\right) \frac{8
    \pi}{\Lambda} \overline{\eta}_\text{GN}^2 \mathcal{Y}_p^2
  (\overline{\eta}_\text{GN}).
    \label{FEnergGNMuZeroBar}
\end{eqnarray}

\begin{figure}[!htb]
\centering \includegraphics[width=8cm]{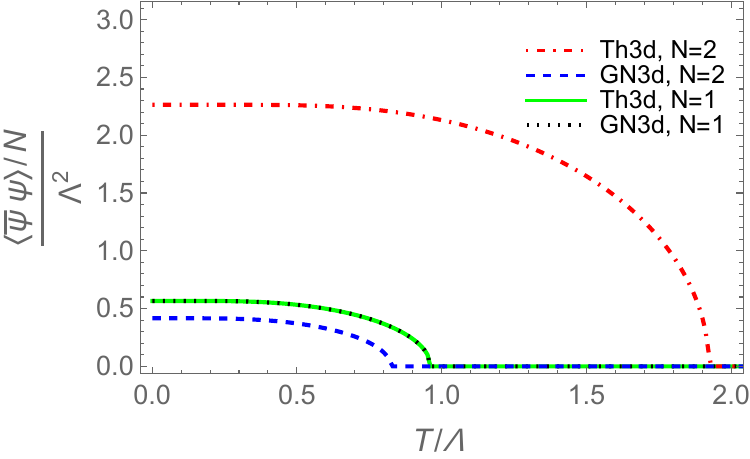}
\caption{The chiral condensate as a function of $T$ for $\mu=0$, when
  $N=1$ and $N=2$.}
\label{FigCondMuZero}
\end{figure}

We can now compare the results obtained for $T_\text{c}$ and $\Omega$
as functions of $N$. Let us start with the critical temperature
difference, $\Delta T_\text{c} =
T_\text{c}^\text{(Th)} - T_\text{c}^\text{(GN)}$, which can be written
as
\begin{equation}
   \Delta T_\text{c} = \frac{4 N}{3} \frac{\Lambda}{2 \ln 2}
   \left(1-\frac{3}{4N-1} \right)\,,
\end{equation}
indicating that the critical temperature for the chiral transition,
predicted by both models, is exactly the same when $N=1$. Moreover,
at this same value of  $N$, Eqs.~(\ref{EtaBarVsTGN}) and
(\ref{EtaBarVsTTh}) imply that $\overline{\eta}_\text{GN} =
\overline{\eta}_\text{Th}$. In this case, using
Eqs.~(\ref{FEnergGNMuZeroBar}) and (\ref{FEnergThMuZeroBar}) one
obtains
\begin{equation}
    \frac{\Delta \Omega(\mu=0, T)}{N} \Bigg|_{N = 1}= 0 \,,
\end{equation}  
which explicitly verifies the predicted equivalence between the models
also when $\mu=0$ and $T\ne 0$.  {}Finally, to get further insights,
we can plot the quark condensate and observe how it is affected by 
temperature variations when some representative values of $N$ are considered. This is explicitly shown in
{}Fig.~\ref{FigCondMuZero}. In both cases, one observes that $\langle \bar{\psi}_k \psi_k \rangle$ decreases in a continuous fashion, suggesting a phase transition of the second kind.

%%%%%%%%%%%%%%%%%%%%%%%%%%%%%%%%%%%%%%%%%%%%%%%%%
%%%%%%%%%%%%%%%%%%%%%%%%%%%%%%%%%%%%%%%%%%%%%%%%%

{}From {}Fig.~\ref{FigCondMuZero}, we can also see that as $N$ decreases,
the fermion condensate grows for the GN model, while for the Thirring
model it moves in the opposite direction, up to $N=1$, when both
condensates finally merge.

%%%%%%%%%%%%%%%%%%%%%%%%%%%%%%%%%%%%%%%%%%%%%%%%%
\subsection{The $T=0$ and $\mu \neq 0$ case}

{}From Eq.~(\ref{Th3dPot}), the effective potential for the Thirring
model in the limit  $T\to 0$ is given by
\begin{eqnarray}
\lefteqn{ \frac{V_{\rm eff}^\text{Th} \left(\eta, {\overline v}_0,
    \mu, T = 0\right)}{N} =   \frac{\overline{v}_0^2 \Lambda}{2
    \pi}-\frac{2\eta^3}{3 \pi }+\frac{3 \eta^4 }{8 \pi N \Lambda}  }
\nonumber \\ &&-\left\{
\frac{1}{2\pi}\left[\frac{(|\eta|-|\mu|)^2}{3}(2
  |\eta|+|\mu|)-2|\eta|^3+ 2\eta^2 |\mu|  \right. \right.  \nonumber
  \\ && \left. \left.- \overline{v}_0\left(\mu^2-\eta^2\right)\right]
\right.  \nonumber \\ &&\left.  + \frac{ 1 }{32 \pi N \Lambda} \left [
  13\eta^4- 14 \eta^2 \mu^2+\mu^4 \right] \right\}
\theta(|\mu|-|\eta|).  \nonumber \\
\label{FreeThT0}
\end{eqnarray}
In this case, the zeroth component of the auxiliary vector field is given by
\begin{equation}
    \overline{v}_0= -\frac{\mu ^2 - \eta ^2}{2 \Lambda }
    \theta(|\mu|-|\eta|).
\end{equation}
At the same time, the GN effective potential, from
Eq. ~(\ref{GN3dPot}) in the $T=0$ limit, reads 
\begin{equation}
\begin{aligned}
\frac{V_{\rm eff}^{\rm GN}\left(\eta, {\overline \sigma}, \mu,
  T=0\right)}{N} &= - \frac{{\overline \sigma}^2 \Lambda}{2 \pi} -
\frac{2 |\eta|^3}{3 \pi} + \frac{ {\overline \sigma}   \eta^2 }{\pi} -
\frac{\eta^4}{8 \pi N \Lambda} \\ &\quad \hspace{-3.5cm}- \Bigg\{
\frac{1}{2 \pi} \left[  \frac{(|\eta| - |\mu| )^2}{3}  (2 |\eta| +
  |\mu| ) + 2 \eta \left(\eta - {\overline \sigma}\right) (|\mu| -
  |\eta|)  \right]  \\ &\quad \hspace{-3.cm} + \frac{1}{32 \pi N
  \Lambda}\left[ 4 \eta^2 \left( \mu^2 - \eta^2 \right) + \left(
  \eta^2 - \mu^2 \right)^2 \right] \Bigg\} \theta(|\mu| - |\eta|)\,,
\label{FreeGNT0}
\end{aligned}
\end{equation}
where
\begin{equation}
{\overline \sigma} = \frac {\eta^2}{\Lambda} + \frac
{\eta}{\Lambda}(|\mu| - |\eta|) \theta(|\mu| - |\eta|) \,.
\end{equation}

%%%%%%%%%%%%%%%%%%%%%%%%%%%%%%%%%%%%%%%%%%%%%%%%%%%%%%%%%%%%%%%%%%%%%%%%%%%%%%
\begin{figure}[!htb]
\centering \includegraphics[width=8cm]{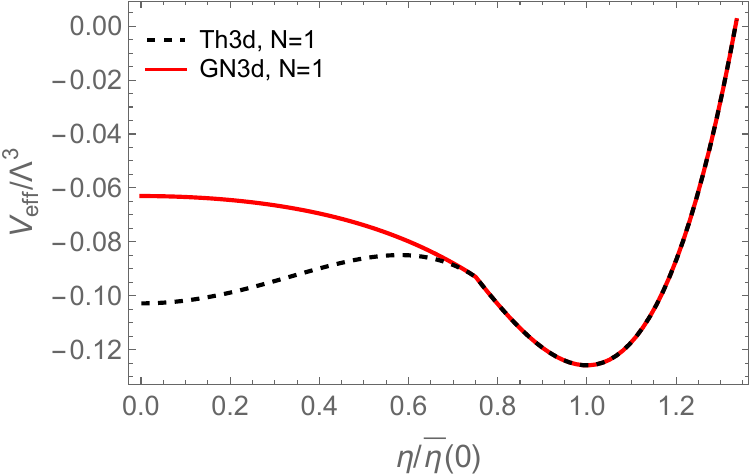}
\caption{The effective potential as a function of $\eta$ for $T=0$ and
  $\mu=\Lambda$, with $N=1$. The variational mass is normalized by
  $\bar{\eta}(0)$, which represents   the optimal value in 
  vacuum (broken) phase, $\bar{\eta}(0)=4\Lambda/3$.}
\label{FigVeffMuZero}
\end{figure}
%%%%%%%%%%%%%%%%%%%%%%%%%%%%%%%%%%%%%%%%%%%%%%%%%%%%%%%%%%%%%%%%%%%%%%%%%%%%%%

As it is well-known, at large-$N$ the GN model suffers a chiral
first-order phase transition when the two phases (symmetric and
broken) coexist at the (coexistence) chemical potential $\mu_\text{c}
= \Lambda$. Going beyond the large-$N$ approximation, it has been
shown that the value of $\mu_\text{c}$ increases as $N$
decreases~\cite{Kneur:2007vj,Kneur:2007vm}. To review how
$\mu_\text{c}$ can be determined, let us examine both effective
potentials. In {}Fig.~\ref{FigVeffMuZero} we show this quantity for each model,  
at $N=1$,  as a function of the variational masses for the
emblematic (GN large-$N$) value  $\mu_c = \Lambda$. The figure illustrates
part of a typical first-order transition pattern. At this
chemical potential value the global (stable) minima for both models
lies at ${\overline \eta}_\text{GN} = {\overline \eta}_\text{Th}
\equiv {\overline \eta} (0)$, where  ${\overline \eta} (0) = 4 \Lambda
/3$ represents the optimal variational mass corresponding to the
vacuum (broken) phase for the relevant $N=1$ case (see Eqs.~(\ref {EtaBarMuTZeroTh})
and (\ref {EtaBarMuTZeroGN})). The dynamics of the first-order phase
transition can then be understood by observing how the effective
potential extremum, at ${\overline \eta}_\text{GN} = {\overline
  \eta}_\text{Th}  \equiv 0$, behave as $\mu \to \mu_\text{c}$. When
$\mu=0$, $V_{\rm eff}$ is characterized by a maximum at the
origin and then, as $\mu$ further increases, this maximum becomes an
inflection point signaling the first spinodal. At even higher
$\mu$ values the inflection becomes a local (metastable) minimum, as 
{}Fig.~\ref{FigVeffMuZero} suggests. As discussed in
Refs.~\cite{Kneur:2007vj,Kneur:2007vm} for the GN case, the phase
transition happens when the minima at the origin and at ${\overline
  \eta} (0)$ become degenerate. In this case, the chemical potential
value at which the broken (vacuum) and symmetric (compressed) matter
coexist can be determined from $V_{\rm
  eff}(\overline{\eta}(0),\mu_\text{c}) = V_{\rm
  eff}(\overline{\eta}=0,\mu_\text{c})$. Recalling that  $V_{\rm
  eff}(\overline{\eta})=\Omega$ one can determine $\mu_\text{c}$ in
the Thirring case by equating Eq.~(\ref{FEnergThTMuZeroBar}) to   
\begin{equation}
\begin{split}
    \Omega_\text{Th} \left(\overline{\eta} = 0, \overline{v}_0, \mu, T
    = 0\right) & = \frac{\overline{v}_0^2 \Lambda}{2 \pi} -
    \frac{\overline{v}_0 \mu^2}{2 \pi} - \frac{|\mu|^3}{6 \pi} -
    \frac{\mu^4}{32 \pi N \Lambda} \\ &\hspace{-2.cm} = -\frac{|\mu|
      ^3 }{6 \pi } \left[1+\left(1 + \frac{1}{4 N} \right)\frac{3
        |\mu| }{4  \Lambda }\right] \, ,
\end{split}
\label{OmegaThEtaBarZero}
\end{equation}
obtaining
\begin{equation}
\big|\mu_{\text{c}}^\text{(Th)}\big| = \frac{4 \Lambda N}{3
}\left[1+\left( 1 + \frac{1}{4N} \right) \frac{3  }{4 \Lambda}
  \big|\mu_{\text{c}}^\text{(Th)}\big|\right]^{-1 / 3}\,.
\label{MUtilCrit}
\end{equation}
Adopting the same strategy in the GN case, one must equate Eq.~(\ref
{FEnergGNTMuZeroBar}) to
\begin{equation}
\Omega_\text{GN} \left(\overline{\eta} = 0, \overline{\sigma}=0, \mu,
T = 0\right)  = -\frac{|\mu| ^3 }{6 \pi } \left(1+\frac{3}{16 N}
\frac{\left|\mu\right|}{\Lambda}\right)\,,
\label{OmegaGNEtaBarZero}
\end{equation}
which gives the result
\begin{equation}
\big|\mu_\text{c}^\text{(GN)}\big| = \frac{ \Lambda}{1 -
  \frac{1}{4N}}\left(1+\frac{3}{16 N}
\frac{\big|\mu_\text{c}^\text{(GN)}\big|}{\Lambda}\right)^{-1 / 3}\,,
\end{equation}
and which confirms the well-established large-$N$ result, $\mu_c=\Lambda$,  when $N \to \infty$~ \cite{Rosenstein:1990nm}.
Then, defining $\Delta \left|\mu_{\text{c}}\right| =
\big|\mu_{\text{c}}^\text{(Th)}\big| -
\big|\mu_{\text{c}}^\text{(GN)}\big|$\,, 
allows us to write
\begin{align}
\Delta \left|\mu_{\text{c}}\right| &= \frac{4\Lambda N}{3} \left\{
  \frac{1}{\left[1 + \left(1+\frac{1}{4N} \right) \frac{3}{4 \Lambda}
    \big|\mu_{\text{c}}^\text{(Th)}\big| \right]^{1/3}} \right. \notag
  \\ &\quad \left. - \frac{3}{\left(4N - 1\right) \left(1 + \frac{3
    }{16 \Lambda N}
    \big|\mu_{\text{c}}^\text{(GN)}\big|\right)^{1/3}} \right\}.
\end{align}

%%%%%%%%%%%%%%%%%%%%%%%%%%%%%%%%%%%%%%%%%%%%%%%%%%%%%%%%%%%%%%%%%%%%%%%%%%%%%%
\begin{figure}[!htb]
\centering \includegraphics[width=8cm]{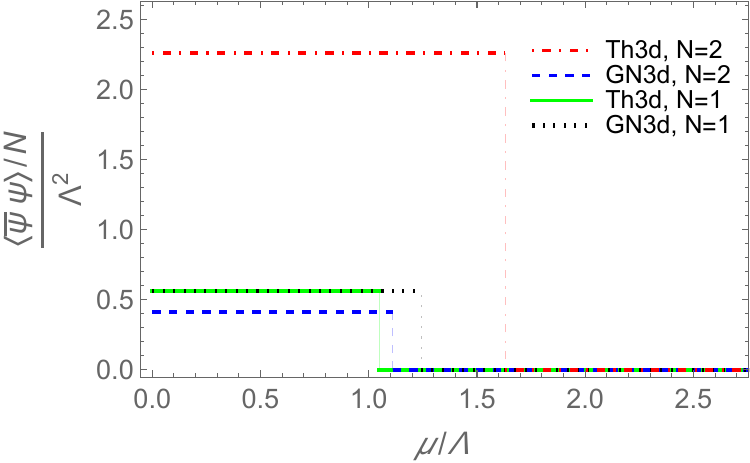}
\caption{The fermion condensate as function of $\mu$ for $T=0$ when
  $N=1$ and $N=2$.}
\label{fig4}
\end{figure}
%%%%%%%%%%%%%%%%%%%%%%%%%%%%%%%%%%%%%%%%%%%%%%%%%%%%%%%%%%%%%%%%%%%%%%%%%%%%%%

As one could anticipate from {}Fig.~\ref{fig4}, which shows the chiral
fermion condensate for both the GN and Thirring models as a function
of the chemical potential the two models do not
predict an equivalent result for $\mu_\text{c}$ even when $N=1$, at
least in the first order in the OPT scheme, since
\begin{align}
\Delta \left|\mu_{\text{c}}\right| &= \frac{4 \Lambda}{3} \left[
  \left(1+\frac{15}{16 \Lambda}
  \big|\mu_{\text{c}}^\text{(Th)}\big|\right)^{-1 / 3} \right. \notag
  \\ &\quad \left. - \left(1+\frac{3}{16 \Lambda}
  \big|\mu_{\text{c}}^\text{(GN)}\big|\right)^{-1 / 3} \right]\,.
\label{Deltamuc}
\end{align}
Indeed, for $N=1$, an explicit evaluation gives
$\big|\mu_{\text{c}}^\text{(Th)}\big| = 1.05946 \Lambda$ and
$\big|\mu_\text{c}^\text{(GN)}\big|  1.24337 \Lambda$, such that $\Delta
\left|\mu_{\text{c}}\right|= 0.18391 \Lambda$, as 
Eq.~(\ref{Deltamuc}) predicts.  To further understand the origin of such discrepancy,
we  can now check the equivalence of the thermodynamic potentials at
the two degenerate minima, starting with the one which represents the
broken (vacuum) phase. In this case, Eqs.~(\ref{FEnergThTMuZeroBar})
and (\ref{FEnergGNTMuZeroBar}) explicitly show  that both
thermodynamic potentials exactly coincide at ${\overline \eta}(0)$, in
agreement with the result shown in {}Fig.~\ref{FigVeffMuZero}.  {}
On the other hand,  the difference between the thermodynamic
potentials at the origin as given by Eqs.~(\ref{OmegaThEtaBarZero})
and (\ref{OmegaGNEtaBarZero}) gives the following nonvanishing result
\begin{equation}
    \frac{\Delta \Omega({\overline \eta}=0, \mu, T=0)}{N} \Bigg|_{N =
      1} = -\frac{  \mu^4}{8 \pi \Lambda} = - \frac{\overline{v}_0(\overline{\eta}_\text{Th}=0)^2
      \Lambda }{2 \pi}\,,
    \label{FEnergGNMuZeroBar2}
\end{equation}
which,  precisely agrees with Eq. (\ref {equivalence}) upon setting $\eta=0$ in the latter. 
{}From Eq.~(\ref{FEnergGNMuZeroBar2}), we can confirm that, as observed in Sec. \ref{section4},  the
agreement between the GN and the Thirring models does not hold at finite densities. This difference is a result of the vector field expectation value contribution to the Thirring model, which does not seem to have a counterpart within the GN model,  at least at the approximation level adopted here. In this context, it is worth recalling that the OPT (radiatively)
generates a vectorlike contribution to the GN model through the two-loop term proportional to $(\delta\lambda/N) \mathcal{W}_p^2$
~\cite{Kneur:2012qp,Restrepo:2014fna}. However, this term is not sufficient to compensate the similar type of contributions arising within the Thirring model at $N=1$.

%%%%%%%%%%%%%%%%%%%%%%%%%%%%%%%%%%%%%%%%%%%%%%%%%%%%%%%%%%%%%%%%%%%%%%%%%%%%%%
\begin{figure}[!htb]
\centering \includegraphics[width=8cm]{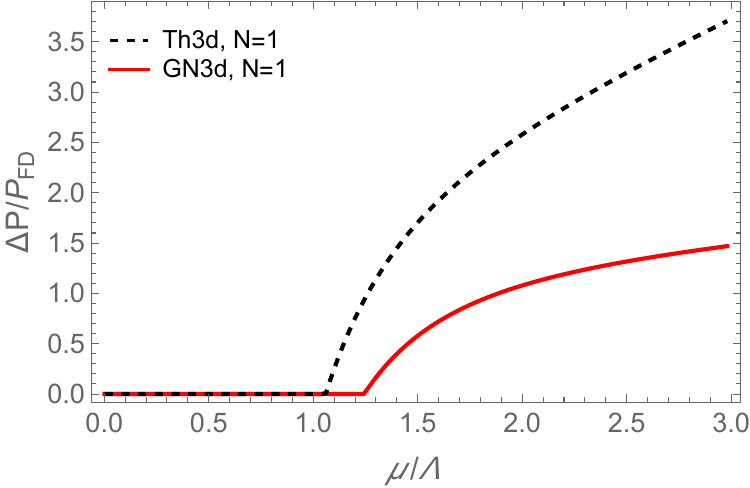}
\caption{The normalized pressure difference for the Thirring and GN models at $N=1$ and
  $T=0$ as function of the chemical potential.  $P_{FD} = \mu^3/(6\pi)$ is the free Fermi-Dirac gas pressure for massless fermions.}
\label{fig5}
\end{figure}
%%%%%%%%%%%%%%%%%%%%%%%%%%%%%%%%%%%%%%%%%%%%%%%%%

As a consequence of the discussion given above, the observed discrepancy also manifests itself in other thermodynamic
quantities as the chemical potential increases. As an example, in
{}Fig.~\ref{fig5} we show the pressure for both models at $N=1$ and
$T=0$ as a function of the chemical potential. In this figure, where the pressure is normalized by the pressure of the free Fermi-Dirac gas of massless fermions, $P_{FD} = \mu^3/(6\pi)$, one can see how the pressure difference rises with increasing values of $\mu$.

%%%%%%%%%%%%%%%%%%%%%%%%%%%%%%%%%%%%%%%%%%%%%%%%%
\subsection{The $T\neq 0$ and $\mu \neq 0$ case}

%%%%%%%%%%%%%%%%%%%%%%%%%%%%%%%%%%%%%%%%%%%%%%%%%%%%%%
\begin{center}
\begin{figure}[!htpb]
\subfigure[]{\includegraphics[width=8cm]{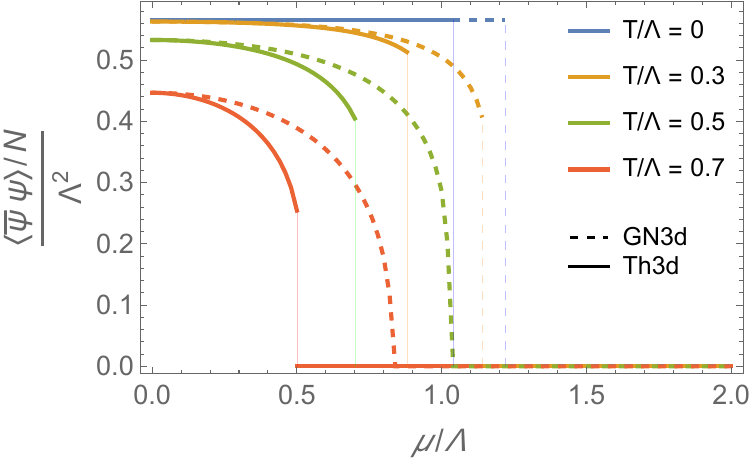}}
\subfigure[]{\includegraphics[width=8cm]{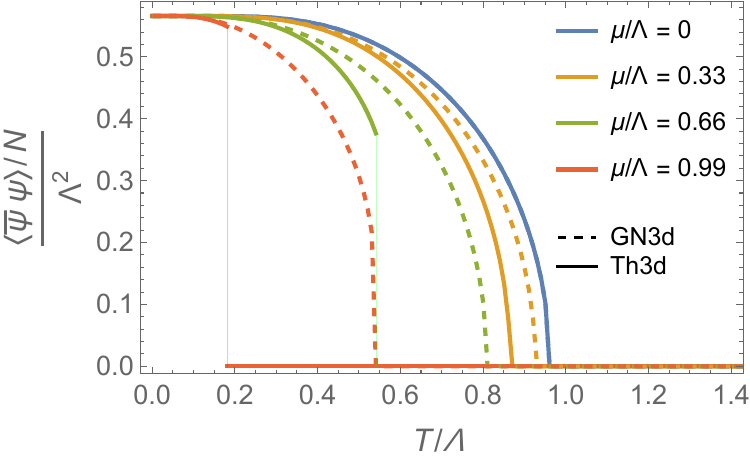} }
\caption{The fermion condensate for the Thirring and GN models at
  $N=1$, as a function of the chemical potential, for fixed values of
  the temperature (panel a) and as a function of the temperature, for
  fixed values of the chemical potential (panel b).}
\label{fig6}
\end{figure}
\end{center}
%%%%%%%%%%%%%%%%%%%%%%%%%%%%%%%%%%%%%%%%%%%%%%%%%%%%%%

{}For completeness and illustration purposes, let us also (numerically) study the case relevant for a hot and dense system of fermions starting with  {}Fig.~\ref{fig6}
which displays the fermion condensates, as a function
of $\mu$, for fixed values of the temperature (panel
a) and as a function of the temperature, for fixed values of $\mu$ (panel b). Note that as the chemical potential
increases, the discrepancy between each model increases, as expected
from all the previous discussions.

%%%%%%%%%%%%%%%%%%%%%%%%%%%%%%%%%%%%%%%%%%%%%%%%%%%%%%
\begin{center}
\begin{figure}[!htpb]
\includegraphics[width=8cm]{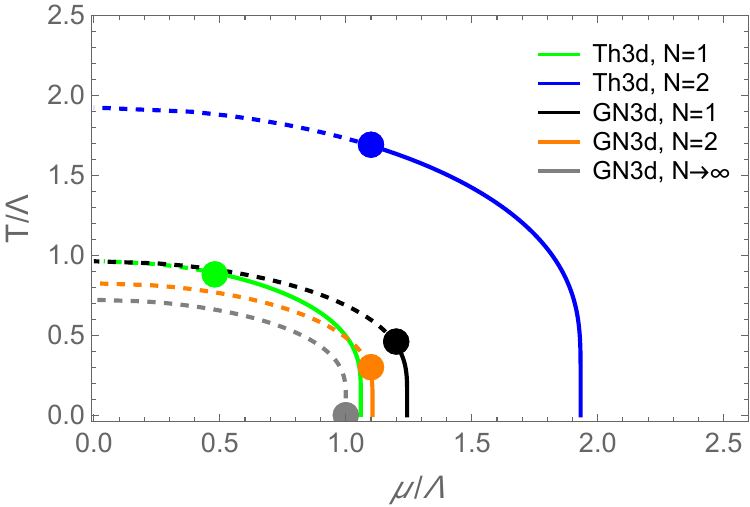}
\caption{The phase diagram for the GN and Thirring models for
  different values of $N$. The dashed lines represent a second-order
  transition boundary, while the solid lines represent the first-order
  transition boundary. The dots represent the tricritical points separating
   the two different types of transitions.}
\label{fig7}
\end{figure}
\end{center}
%%%%%%%%%%%%%%%%%%%%%%%%%%%%%%%%%%%%%%%%%%%%%%%%%%%%%%

{}Finally, in {}Fig.~\ref{fig7} we show the phase diagram, on the $T-\mu$ plane, for each
model at some representative values of $N$.  Once again, it is instructive to  observe the
behavior of each transition boundary as the value of $N$ decreases towards the value
$N=1$. As $N$ decreases, the chiral symmetry broken region for the
Thirring model shrinks, while for the GN model it expands and they seem to
converge to each other.  At $N=1$ they approximately agree to each
other at lower values of the chemical potential, but eventually one
overshoots the other as $\mu$ increases. In particular, the
tricritical points representing each model 
are quite separated, even at $N=1$.  

%%%%%%%%%%%%%%%%%%%%%%%%%%%%%%%%%%%%%%%%%%%%%%%%%
\section{Conclusions}
\label{conclusions}

In this paper,  we have analyzed the duality between the Thirring and
Gross-Neveu models in $2+1$-dimensions in order to explicitly confirm that they are equivalent when $N=1$, in accordance with the predictions from the {}Fierz
identities. As both models are typically studied in the large-$N$
approximation, which is simply not reliable enough to study the
predicted equivalence in the case of a single flavor, here we have made use of
the   optimized perturbation theory approximation which can easily incorporate  $1/N$ contributions. 

{}For our purposes, the OPT capability to generate corrections going beyond the simple
(mean-field) large-$N$ predictions already at the first nontrivial order is of particular interest, since it offers a unique twofold opportunity: gauge the method's reliability and
check the predicted equivalence between the two models. {}Considering 
the two distinct effective potentials, at 
 first perturbative order, we have then compared the resulting phase transition
patterns at finite $N$ values. As far as the
duality regime at $N=1$ is concerned,  we have demonstrated that it indeed holds
whenever the density vanishes. In such
cases, the OPT at its first nontrivial order has shown to be
already quite appropriate. At finite values of $\mu$,
however, the equivalence of both models for $N=1$ could not be fully confirmed, at least at the approximation level considered here. Yet, the approximation was 
sufficient to show that as $N$ decreases from  large-$N$ values
toward  $N=1$, both models display opposing behaviors regarding the
chiral phase transition. While the chiral symmetry
breaking region increases for the GN model as $N$ decreases, the
Thirring model displays an opposite behavior, with the chiral symmetry
breaking region collapsing. In particular, the critical temperatures
for chiral symmetry restoration  satisfy $T_c^{(\rm
  GN)} < T_c^{(\rm Thirring)}$ for $N>1$ and $T_c^{(\rm
  GN)} = T_c^{(\rm Thirring)}$ at
$N=1$. {}For the coexistence chemical potentials we also have a similar
behavior, with  $\mu_c^{(\rm GN)} < \mu_c^{(\rm
  Thirring)}$ when $N>1$, but eventually, at $N=1$, one overshoots the
other. We have traced the latter unexpected discrepancy as a consequence of the
presence of a nonvanishing vacuum expectation value associated with the auxiliary vector
field present in the Thirring model. While it is known that the OPT
approximation can radiatively generate vector
contributions already at its first perturbative order  \cite{Kneur:2012qp,Restrepo:2014fna}, here we have observed that such ($1/N$) contributions to the GN model, proportional to $(\lambda/N) \mathcal{W}_p^2 \sim (\lambda/N) \langle
\psi^{\dagger} \psi \rangle^2 $, are  not enough to
fully compensate all the terms proportional to $\mathcal{W}_p^2$ in the Thirring model (which include $N^0$ and $1/N$ powers). 

The predicted
equivalence between the Thirring and Gross-Neveu models at $N=1$
seems to offer a distinctive opportunity to test different nonperturbative methods
and to gauge their reliability by providing an appropriate, and rather simple, platform.
{}Finally, we remark that this first OPT application to the planar Thirring model predicts that the critical flavor number preventing chiral symmetry breaking for this model is $N=\infty$, in agreement with the  Gaussian approximation results~\cite{Hyun:1994fb}.

In future works, it will be of interest to investigate whether, by
pushing the OPT approximation to higher orders, the differences observed here at finite chemical potentials will eventually cancel, leading to the expected
equivalence when $N=1$. It would also be useful to
explore other nonperturbative methods, capable to go beyond the large-$N$
approximation, to study this same problem.  
In addition to the relation between the Thirring and Gross-Neveu models studied here, we also recall that there are other models that are related to each other
through duality. This includes the equivalence between the Thirring model with the sine–Gordon, the Gross-Neveu-Yukawa, the $U (N )$ massless quantum eletrodynamics models, just to cite a few examples (for a review, see e.g. Ref.~\cite{Moshe:2003xn}). It would be of interest to make the type of study we have done in the present paper to further investigate the relations among those other different models.

%%%%%%%%%%%%%%%%%%%%%%%%%%%%%%%%%%%%%%%%%%%%%%%%%
\begin{acknowledgements}

E.M. is supported by a PhD grant from Conselho Nacional de
Desenvolvimento Cient\'{\i}fico e Tecnol\'{o}gico (CNPq).  Y.M.P.G. is
supported by a postdoctoral grant from Funda\c c\~{a}o Carlos Chagas
Filho de Amparo \`a Pesquisa do Estado do Rio de Janeiro (FAPERJ),
Grant No. E26/201.937/2020.  M.B.P.  is partially supported by
Conselho Nacional de Desenvolvimento Cient\'{\i}fico e Tecnol\'{o}gico
(CNPq), Grant No  307261/2021-2  and by CAPES - Finance  Code  001.
R.O.R. is also partially supported by research grants from Conselho
Nacional de Desenvolvimento Cient\'{\i}fico e Tecnol\'ogico (CNPq),
Grant No. 307286/2021-5, and from {}Funda\c{c}\~ao Carlos Chagas Filho
de Amparo \`a Pesquisa do Estado do Rio de Janeiro (FAPERJ), Grant
No. E-26/201.150/2021. This work has also been financed  in  part  by
Instituto  Nacional  de Ci\^encia  e Tecnologia de F\'{\i}sica Nuclear
e Aplica\c c\~{o}es (INCT-FNA), Process No.  464898/2014-5. 

\end{acknowledgements}

%%%%%%%%%%%%%%%%%%%%%%%%%%%%%%%%%%%%%%%%%%%%%%%%%
\appendix
\section{Dirac matrices trace properties in $2+1$ Dimensions}
\label{AppendixA}

The following expressions involve traces of products of Dirac matrices
in $2+1$ dimensions and which are found in the computation  of the
effective potentials:
\begin{equation}
\begin{gathered}
\operatorname{Tr}\left[ (\eta - \slashed{v})(\slashed{p} + \eta)
  \right] = 4 \left( \eta^2 - p \cdot v \right),
\\ \operatorname{Tr}\left[ (\slashed{p} + \eta)\gamma_\mu (\slashed{q}
  + \eta)\gamma^\mu \right] = 4 \left( 3\eta^2 - p \cdot q \right).
\end{gathered}
\end{equation}
The $\gamma^\mu$ matrices in $2+1$ dimensions are defined as
\begin{equation}
\gamma^\mu = \left(\begin{array}{cc} 1 & 0 \\ 0 & -1
\end{array}\right) \otimes \sigma^\mu = \left(\begin{array}{cc}
\sigma^\mu & 0 \\ 0 & -\sigma^\mu
\end{array}\right),
\end{equation}
with $\sigma^\mu = \left(\sigma^3, i \sigma^1, i \sigma^2\right)$, and
$\bar{\psi} = \psi^\dagger \gamma^0$ is the adjoint spinor. The
$\gamma^\mu$ matrices satisfy the Clifford algebra $\left\{\gamma^\mu,
\gamma^\nu\right\} = 2 g^{\mu \nu} \mathbbm{1}$. They also obey the
identity
\begin{equation}
\gamma^\mu \gamma^\nu = g^{\mu \nu} \mathbbm{1}_{4 \times 4} + i \epsilon^{\mu \nu
  \lambda} \gamma_3 \gamma_\lambda,
\end{equation}
where $\mathbbm{1} = \mathbbm{1}_{2 \times 2}$, $\gamma_3 =
\left(\begin{array}{cc} \mathbbm{1} & 0 \\ 0 &
  -\mathbbm{1}
\end{array}\right)$ and $i \gamma_5 =
\left(\begin{array}{cc} 0 & \mathbbm{1}\\ -\mathbbm{1} &
  0\end{array}\right)$.

The nontrivial traces involving these $\gamma^\mu$ matrices are given
by
\begin{equation}
\begin{aligned}
\operatorname{Tr}\left(\gamma^\mu \gamma^\nu\right) &= 4 g^{\mu \nu},
\\ \operatorname{Tr}\left(\gamma^\mu \gamma^\nu \gamma^\lambda
\gamma_3\right) &= 4 i \epsilon^{\mu \nu \lambda},
\\ \operatorname{Tr}\left(\gamma^\mu \gamma^\nu \gamma^\alpha
\gamma^\beta\right) &=  4 \left(g^{\mu \nu} g^{\alpha \beta} + g^{\mu
  \alpha} g^{\nu \beta} - g^{\mu \beta} g^{\nu \alpha}\right).
\end{aligned}
\end{equation}
The Dirac matrices also satisfy the following identities,
\begin{equation}
\begin{aligned}
\gamma_\mu \gamma^\mu &= 3 \mathbbm{1}, \quad \gamma^\mu \gamma_\nu
\gamma_\mu = -\gamma_\nu, \\ \gamma^\mu \gamma_\nu \gamma_\lambda
\gamma_\mu &= 3 g_{\nu \lambda} \mathbbm{1} - i \epsilon_{\nu \lambda
  \kappa} \gamma^\kappa \gamma_3, \\ \epsilon^{\mu \nu \kappa}
\gamma_\mu \gamma_\nu &= -2 i \gamma^\kappa \gamma_3, \\ \epsilon^{\mu
  \nu \kappa} \gamma_\mu \slashed{a} \gamma_\nu &= -2 i a^\kappa
\gamma_3.
\end{aligned}
\end{equation}

%%%%%%%%%%%%%%%%%%%%%%%%%%%%%%%%%%%%%%%%%%%%%%%%%
\section{Momentum integrals and Matsubara's sums}
\label{AppendixB}

The integrals in Feynman diagrams at finite temperature and density
use the substitution $ p_0 \rightarrow i\left(\omega_n - i \mu\right)
= (2n - 1) i \pi T + \mu $, where $ \mu $ is the chemical potential
and $ \omega_n = (2n + 1) \pi T $ are the Matsubara frequencies for
fermions, with $ n \in \mathbbm{Z} $:
\begin{eqnarray}
&&\int \frac{d^d p}{(2 \pi)^d} f\left(p_0,|\mathbf{p}|\right)
  \nonumber \\ &&\rightarrow i T \sum_{n=-\infty}^{+\infty} \int
  \frac{d^{d-1} \mathbf{p}}{(2 \pi)^{d-1}} f\left[i\left(\omega_n - i
    \mu\right),|\mathbf{p}|\right].
\label{momint}
\end{eqnarray}
Defining  the dispersion relation
\begin{equation}
    E^2_{\mathbf{p}} = \mathbf{p}^2 + \eta^2 \,,
\end{equation}
one can perform the relevant thermal integrals as follows.  Starting
with 

\begin{eqnarray}
\mathcal{X}_p(\eta) &=& i \int \frac{d^d p}{(2 \pi)^d} \ln \left(p^2 -
\eta^2\right)  \nonumber \\ &=& - T \sum_{n=-\infty}^{+\infty} \int
\frac{d^{d-1} \mathbf{p}}{(2 \pi)^{d-1}}  \ln \left[\left(\omega_n - i
  \mu\right)^2 + E^2_{\mathbf{p}}\right], \nonumber \\
\end{eqnarray}
and upon using the Matsubara sum relation one obtains
\begin{eqnarray}
&& T \sum_{n=-\infty}^{+\infty} \ln \left[\left(\omega_n - i
    \mu\right)^2  + E^2_{\mathbf{p}}\right] = E_{\mathbf{p}}
  \nonumber \\ &&+ T \ln \left[1 + e^{-(E_{\mathbf{p}} + \mu) /
      T}\right]  \nonumber \\ &&+ T \ln \left[1 + e^{-(E_{\mathbf{p}}
      - \mu) / T}\right].
\end{eqnarray}
Then, performing the momentum integration within the $\overline {\rm
  MS}$ regularization scheme one obtains
\begin{eqnarray}
\mathcal{X}_p(\eta) &=& \frac{|\eta|^3}{6 \pi} + \frac{\eta}{2 \pi}
T^2  \left\{\operatorname{Li}_2\left[-e^{-(|\eta| - |\mu|) / T}\right]
\right.  \nonumber \\ &+& \left. \operatorname{Li}_2\left[-e^{-(|\eta|
    + |\mu|) / T}\right]\right\} \nonumber \\ & + & \frac{T^3}{2 \pi}
\left\{\operatorname{Li}_3\left[-e^{-(|\eta| - |\mu|) / T}\right]
\right.  \nonumber \\ &+& \left. \operatorname{Li}_3\left[-e^{-(|\eta|
    + |\mu|) / T}\right]\right\}.  \nonumber \\
\label{XTmu}
\end{eqnarray}

Next, let us consider the integral
\begin{eqnarray}
\mathcal{Y}_p(\eta) &=& i \int \frac{d^d p}{(2 \pi)^d} \frac{1}{p^2 -
  \eta^2}  \nonumber \\ &=& T \sum_{n=-\infty}^{+\infty} \int
\frac{d^{d-1} \mathbf{p}}{(2 \pi)^{d-1}}  \frac{1}{\left(\omega_{n} -
  i \mu\right)^2 + E_{\mathbf{p}}^2}.
\end{eqnarray}
In this case,  the Matsubara sum reads
\begin{eqnarray}
&&T \sum_{n=-\infty}^{+\infty} \frac{1}{\left(\omega_n - i
    \mu\right)^2 + E_{\mathbf{p}}^2}  \nonumber\\ &&= \frac{1}{2
    E_{\mathbf{p}}}\left[1 - \frac{1}{e^{(E_{\mathbf{p}} + \mu) / T} +
      1} - \frac{1}{e^{(E_{\mathbf{p}} - \mu) / T} + 1}\right] \;,
  \nonumber \\
\end{eqnarray}
and performing the momentum integration, one obtains
\begin{eqnarray}
\mathcal{Y}_p(\eta) &=& -\frac{T}{4 \pi}\left\{\frac{|\eta|}{T}
\right.  \nonumber \\ &+& \left. \ln \left[1 + e^{-(|\eta| - |\mu|) /
    T}\right] + \ln \left[1 + e^{-(|\eta| + |\mu|) /
    T}\right]\right\}.  \nonumber \\
\label{YTmu}
\end{eqnarray}

{}Finally, let us consider 
\begin{eqnarray}
\mathcal{W}_p(\eta) &=& i \int \frac{d^d p}{(2 \pi)^d} \frac{p_0}{p^2
  - \eta^2}  \nonumber \\ &=& i T \sum_{n=-\infty}^{+\infty} \int
\frac{d^{d-1} p}{(2 \pi)^{d-1}}  \frac{\left(\omega_n - i
  \mu\right)}{\left(\omega_n - i \mu\right)^2 + E_{\mathbf{p}}^2},
\nonumber \\
\end{eqnarray}
and  the Matsubara sum relation
\begin{eqnarray}
&&T \sum_{n=-\infty}^{+\infty} \frac{\left(\omega_n - i
    \mu\right)}{\left(\omega_n - i \mu\right)^2 + E_{\mathbf{p}}^2}
  \nonumber \\ &=& \frac{i}{2}\left[\frac{1}{1 +
      e^{\left(E_{\mathbf{p}} - \mu\right) / T}}  - \frac{1}{1 +
      e^{\left(E_{\mathbf{p}} + \mu\right) / T}}\right],
\end{eqnarray}
which, after integrating over $\mathbf{p}$ gives 
\begin{eqnarray}
\mathcal{W}_p(\eta) &=& -\operatorname{sgn}(\mu) \frac{T^2}{4
  \pi}\left\{\frac{|\eta|}{T}  \ln \left[\frac{1 + e^{(|\eta| + |\mu|)
      / T}}{1 + e^{(|\eta| - |\mu|) / T}}\right]  \right.  \nonumber
\\ &+& \left. \operatorname{Li}_2\left[-e^{(|\eta| + |\mu|) /
    T}\right]  - \operatorname{Li}_2\left[-e^{(|\eta| - |\mu|) /
    T}\right]\right\}, \nonumber \\
\label{WTmu}
\end{eqnarray}
which only contributes at finite densities ($\mu\neq 0$).

It is useful also to quote the  $T \to 0$ limit of the above
functions,
\begin{eqnarray}
\mathcal{X}_p(\eta)\Bigr|_{T\to 0} &=& \frac{|\eta|^3}{6 \pi} -
\frac{\eta}{4 \pi} (|\mu| - |\eta|)^2 \theta(|\mu| - |\eta|)
\nonumber\\ &+& \frac{1}{12 \pi} (|\eta| - |\mu|)^3 \theta(|\mu| -
|\eta|), 
\label{XpTZero}
\end{eqnarray}
\begin{eqnarray}
\mathcal{Y}_p(\eta)\Bigr|_{T\to 0} &=& -\frac{|\eta|}{4 \pi} - \frac{1}{4 \pi} (|\mu|
- |\eta|) \theta(|\mu| - |\eta|), 
\nonumber \\
\label{YpTZero}
\end{eqnarray}
\begin{eqnarray}
\mathcal{W}_p(\eta)\Bigr|_{T\to 0} &=& -\frac{1}{8 \pi} \operatorname{sgn}(\mu)
\left(\mu^2  - \eta^2\right) \theta(|\mu| - |\eta|). 
\nonumber \\
\label{WpTZero}
\end{eqnarray}

%%%%%%%%%%%%%%%%%%%%%%%%%%%%%%%%%%%%%%%%%%%%%%%%%


\begin{thebibliography}{99}


%\cite{Nambu:1961tp}
\bibitem{Nambu:1961tp}
Y.~Nambu and G.~Jona-Lasinio,
``Dynamical Model of Elementary Particles Based on an Analogy with Superconductivity. I.,''
Phys. Rev. \textbf{122}, 345-358 (1961)
\href{https://doi.org/10.1103/PhysRev.122.345}{doi:10.1103/PhysRev.122.345}
%6170 citations counted in INSPIRE as of 03 Jul 2024


%\cite{Nambu:1961fr}
\bibitem{Nambu:1961fr}
Y.~Nambu and G.~Jona-Lasinio,
``Dynamical model of elementary particles based on an analogy with superconductivity. II.,''
Phys. Rev. \textbf{124}, 246-254 (1961)
\href{https://doi.org/10.1103/PhysRev.124.246}{doi:10.1103/PhysRev.124.246}
%3157 citations counted in INSPIRE as of 03 Jul 2024



%\cite{Gross:1974jv}
\bibitem{Gross:1974jv}
D.~J.~Gross and A.~Neveu,
``Dynamical Symmetry Breaking in Asymptotically Free Field Theories,''
Phys. Rev. D \textbf{10}, 3235 (1974)
\href{https://doi.org/10.1103/PhysRevD.10.3235}{doi:10.1103/PhysRevD.10.3235}
%2060 citations counted in INSPIRE as of 27 Jun 2024


%\cite{Thirring:1958in}
\bibitem{Thirring:1958in}
W.~E.~Thirring,
``A Soluble Relativistic Field Theory?,''
Annals Phys. \textbf{3}, 91 (1958)
\href{https://doi.org/10.1016/0003-4916(58)90015-0}{doi:10.1016/0003-4916(58)90015-0}
%629 citations counted in INSPIRE as of 01 Jul 2024


%\cite{Hands:2008id}
\bibitem{Hands:2008id}
S.~Hands and C.~Strouthos,
``Quantum Critical Behaviour in a Graphene-like Model,''
Phys. Rev. B \textbf{78}, 165423 (2008)
\href{https://doi.org/10.1103/PhysRevB.78.165423}{doi:10.1103/PhysRevB.78.165423}
\href{https://arxiv.org/pdf/0806.4877}{[arXiv:0806.4877 [cond-mat.str-el]]}.
%104 citations counted in INSPIRE as of 29 Feb 2024


%\cite{Ebert:2015hva}
\bibitem{Ebert:2015hva}
D.~Ebert, K.~G.~Klimenko, P.~B.~Kolmakov and V.~C.~Zhukovsky,
``Phase Transitions in Hexagonal, Graphene-Like Lattice Sheets and Nanotubes Under the Influence of External Conditions,''
Annals Phys. \textbf{371}, 254 (2016)
\href{https://doi.org/10.1016/j.aop.2016.05.001}{doi:10.1016/j.aop.2016.05.001}
\href{https://arxiv.org/pdf/1509.08093}{[arXiv:1509.08093 [cond-mat.mes-hall]]}.
%27 citations counted in INSPIRE as of 13 Mar 2024


%\cite{Ebert:2018dzs}
\bibitem{Ebert:2018dzs}
D.~Ebert and D.~Blaschke,
``Thermodynamics of a Generalized Graphene-Motivated (2+1) D Gross\textendash{}Neveu Model Beyond the Mean Field Within the Beth\textendash{}Uhlenbeck Approach,''
PTEP \textbf{2019}, 123I01 (2019)
\href{https://doi.org/10.1093/ptep/ptz110}{doi:10.1093/ptep/ptz110}
\href{https://arxiv.org/pdf/1811.07109}{[arXiv:1811.07109 [cond-mat.mes-hall]]}.
%9 citations counted in INSPIRE as of 29 Feb 2024


%\cite{Zhukovsky:2017hzo}
\bibitem{Zhukovsky:2017hzo}
V.~C.~Zhukovsky, K.~G.~Klimenko and T.~G.~Khunjua,
``Superconductivity in Chiral-Asymmetric Matter within the (2+1)-Dimensional Four-Fermion Model,''
Moscow Univ. Phys. Bull. \textbf{72}, 250 (2017)
\href{https://doi.org/10.3103/S002713491703016X}{doi:10.3103/S002713491703016X}
%7 citations counted in INSPIRE as of 29 Feb 2024


%\cite{Klimenko:2012qi}
\bibitem{Klimenko:2012qi}
K.~G.~Klimenko, R.~N.~Zhokhov and V.~C.~Zhukovsky,
``Superconductivity Phenomenon Induced by External In-Plane Magnetic Field in (2+1)-Dimensional Gross-Neveu Type Model,''
Mod. Phys. Lett. A \textbf{28}, 1350096 (2013)
\href{https://doi.org/10.1142/S021773231350096X}{doi:10.1142/S021773231350096X}
\href{https://arxiv.org/pdf/1211.0148}{[arXiv:1211.0148 [hep-th]]}.
%15 citations counted in INSPIRE as of 13 Mar 2024


%\cite{Gomes:2021nem}
\bibitem{Gomes:2021nem}
Y.~M.~P.~Gomes and R.~O.~Ramos,
``Tilted Dirac Cone Effects and Chiral Symmetry Breaking in a Planar Four-Fermion Model,''
Phys. Rev. B \textbf{104}, 245111 (2021)
\href{https://doi.org/10.1103/PhysRevB.104.245111}{doi:10.1103/PhysRevB.104.245111}
\href{https://arxiv.org/pdf/2106.09239}{[arXiv:2106.09239 [cond-mat.mes-hall]]}.
%10 citations counted in INSPIRE as of 06 Jun 2024


%\cite{Gomes:2022dmf}
\bibitem{Gomes:2022dmf}
Y.~M.~P.~Gomes and R.~O.~Ramos,
``Superconducting Phase Transition in Planar Fermionic Models with Dirac Cone Tilting,''
Phys. Rev. B \textbf{107}, 125120 (2023)
\href{https://doi.org/10.1103/PhysRevB.107.125120}{doi:10.1103/PhysRevB.107.125120}
\href{https://arxiv.org/pdf/2204.08534}{[arXiv:2204.08534 [cond-mat.str-el]]}.
%6 citations counted in INSPIRE as of 06 Jun 2024


%\cite{Gomes:2023vvu}
\bibitem{Gomes:2023vvu}
Y.~M.~P.~Gomes, E.~Martins, M.~B.~Pinto and R.~O.~Ramos,
``First-Order Phase Transitions within Weyl Type of Materials at Low Temperatures,''
Phys. Rev. B \textbf{108}, 085107 (2023)
\href{https://doi.org/10.1103/PhysRevB.108.085107}{doi:10.1103/PhysRevB.108.085107}
\href{https://arxiv.org/pdf/2305.09007}{[arXiv:2305.09007 [cond-mat.str-el]]}.
%1 citations counted in INSPIRE as of 29 Feb 2024


%\cite{Caldas:2008zz}
\bibitem{Caldas:2008zz}
H.~Caldas, J.~L.~Kneur, M.~B.~Pinto and R.~O.~Ramos,
``Critical Dopant Concentration in Polyacetylene and Phase Diagram from a Continuous Four-Fermi model,''
Phys. Rev. B \textbf{77}, 205109 (2008)
\href{https://doi.org/10.1103/PhysRevB.77.205109}{doi:10.1103/PhysRevB.77.205109}
\href{https://arxiv.org/pdf/0804.2675}{[arXiv:0804.2675 [cond-mat.soft]]}.
%31 citations counted in INSPIRE as of 29 Feb 2024


%\cite{Caldas:2009zz}
\bibitem{Caldas:2009zz}
H.~Caldas and R.~O.~Ramos,
``Magnetization of Planar Four-Fermion Systems,''
Phys. Rev. B \textbf{80}, 115428 (2009)
\href{https://doi.org/10.1103/PhysRevB.80.115428}{doi:10.1103/PhysRevB.80.115428}
\href{https://arxiv.org/pdf/0907.0723}{[arXiv:0907.0723 [cond-mat.soft]]}.
%32 citations counted in INSPIRE as of 29 Feb 2024


%\cite{Ramos:2013aia}
\bibitem{Ramos:2013aia}
R.~O.~Ramos and P.~H.~A.~Manso,
``Chiral Phase Transition in a Planar Four-Fermi Model in a Tilted Magnetic Field,''
Phys. Rev. D \textbf{87}, 125014 (2013)
\href{https://doi.org/10.1103/PhysRevD.87.125014}{doi:10.1103/PhysRevD.87.125014}
\href{https://arxiv.org/pdf/1303.5463}{[arXiv:1303.5463 [hep-ph]]}.
%15 citations counted in INSPIRE as of 29 Feb 2024


%\cite{Klimenko:2013gua}
\bibitem{Klimenko:2013gua}
K.~G.~Klimenko and R.~N.~Zhokhov,
``Magnetic Catalysis Effect in the (2+1)-Dimensional Gross-Neveu Model With Zeeman Interaction,''
Phys. Rev. D \textbf{88}, 105015 (2013)
\href{https://doi.org/10.1103/PhysRevD.88.105015}{doi:10.1103/PhysRevD.88.105015}
\href{https://arxiv.org/pdf/1307.7265}{[arXiv:1307.7265 [hep-ph]]}.
%15 citations counted in INSPIRE as of 29 Feb 2024


%\cite{Khunjua:2021fus}
\bibitem{Khunjua:2021fus}
T.~G.~Khunjua, K.~G.~Klimenko and R.~N.~Zhokhov,
``Spontaneous Non-Hermiticity in the (2+1)-dimensional Gross-Neveu Model,''
Phys. Rev. D \textbf{105}, 025014 (2022)
\href{https://doi.org/10.1103/PhysRevD.105.025014}{doi:10.1103/PhysRevD.105.025014}
\href{https://arxiv.org/pdf/2112.13012}{[arXiv:2112.13012 [hep-th]]}.
%6 citations counted in INSPIRE as of 29 Feb 2024


%\cite{Gubaeva:2022feb}
\bibitem{Gubaeva:2022feb}
M.~M.~Gubaeva, T.~G.~Khunjua, K.~G.~Klimenko and R.~N.~Zhokhov,
``Spontaneous Non-Hermiticity in the (2+1)-Dimensional Thirring Model,''
Phys. Rev. D \textbf{106}, 125010 (2022)
\href{https://doi.org/10.1103/PhysRevD.106.125010}{doi:10.1103/PhysRevD.106.125010}
\href{https://arxiv.org/pdf/2212.01062}{[arXiv:2212.01062 [hep-th]]}.
%5 citations counted in INSPIRE as of 06 Jun 2024


%\cite{Khunjua:2022kxf}
\bibitem{Khunjua:2022kxf}
T.~G.~Khunjua, K.~G.~Klimenko and R.~N.~Zhokhov,
``Hartree-Fock Approach to Dynamical Mass Generation in the Generalized (2+1)-Dimensional Thirring Model,''
Phys. Rev. D \textbf{106}, 085002 (2022)
\href{https://doi.org/10.1103/PhysRevD.106.085002}{doi:10.1103/PhysRevD.106.085002}
\href{https://arxiv.org/pdf/2208.11735}{[arXiv:2208.11735 [hep-th]]}.
%2 citations counted in INSPIRE as of 06 Jun 2024


%\cite{Fierz:1937wjm}
\bibitem{Fierz:1937wjm}
M.~Fierz,
``Zur Fermischen Theorie des $\beta$-Zerfalls,''
Z. Phys. \textbf{104}, 553 (1937)
\href{https://doi.org/10.1007/bf01330070}{doi:10.1007/bf01330070}
%167 citations counted in INSPIRE as of 04 Jun 2024


%\cite{Nieves:2003in}
\bibitem{Nieves:2003in}
J.~F.~Nieves and P.~B.~Pal,
``Generalized Fierz Identities,''
Am. J. Phys. \textbf{72}, 1100 (2004)
\href{https://doi.org/10.1119/1.1757445}{doi:10.1119/1.1757445}
\href{https://arxiv.org/pdf/hep-ph/0306087}{[arXiv:hep-ph/0306087 [hep-ph]]}.
%105 citations counted in INSPIRE as of 24 Jun 2024


%\cite{Rosenstein:1988zf}
\bibitem{Rosenstein:1988zf}
B.~Rosenstein and A.~Kovner,
``Gross-Neveu and Thirring Models. Covariant Gaussian Analysis,''
Phys. Rev. D \textbf{40}, 523 (1989)
\href{https://doi.org/10.1103/PhysRevD.40.523}{doi:10.1103/PhysRevD.40.523}
%9 citations counted in INSPIRE as of 29 Feb 2024



%\cite{Wipf:2022hqd}
\bibitem{Wipf:2022hqd}
A.~W.~Wipf and J.~J.~Lenz,
``Symmetries of Thirring Models on 3D Lattices,''
Symmetry \textbf{14}, 333 (2022)
\href{https://doi.org/10.3390/sym14020333}{doi:10.3390/sym14020333}
\href{https://arxiv.org/pdf/2201.01692}{[arXiv:2201.01692 [hep-lat]]}.
%5 citations counted in INSPIRE as of 29 Feb 2024


%\cite{Karsch:2001cy}
\bibitem{Karsch:2001cy}
F.~Karsch,
``Lattice QCD at High Temperature and Density,''
Lect. Notes Phys. \textbf{583}, 209 (2002)
\href{https://doi.org/10.1007/3-540-45792-5_6}{doi:10.1007/3-540-45792-5\_6}
\href{https://arxiv.org/pdf/hep-lat/0106019}{[arXiv:hep-lat/0106019 [hep-lat]]}.
%796 citations counted in INSPIRE as of 26 Jun 2024


%\cite{Muroya:2003qs}
\bibitem{Muroya:2003qs}
S.~Muroya, A.~Nakamura, C.~Nonaka and T.~Takaishi,
``Lattice QCD at Finite Density: An Introductory Review,''
Prog. Theor. Phys. \textbf{110}, 615 (2003)
\href{https://doi.org/10.1143/PTP.110.615}{doi:10.1143/PTP.110.615}
\href{https://arxiv.org/pdf/hep-lat/0306031}{[arXiv:hep-lat/0306031 [hep-lat]]}.
%238 citations counted in INSPIRE as of 02 Jul 2024


%\cite{Okopinska:1987hp}
\bibitem{Okopinska:1987hp}
A.~Okopinska,
``Nonstandard Expansion Techniques for the Effective Potential in $\lambda \phi^4$ Quantum Field Theory,''
Phys. Rev. D \textbf{35}, 1835 (1987)
\href{https://doi.org/10.1103/PhysRevD.35.1835}{doi:10.1103/PhysRevD.35.1835}
%198 citations counted in INSPIRE as of 29 Feb 2024


%\cite{Duncan:1988hw}
\bibitem{Duncan:1988hw}
A.~Duncan and M.~Moshe,
``Nonperturbative Physics from Interpolating Actions,''
Phys. Lett. B \textbf{215}, 352 (1988)
\href{https://doi.org/10.1016/0370-2693(88)91447-5}{doi:10.1016/0370-2693(88)91447-5}
%151 citations counted in INSPIRE as of 12 Jun 2024


%\cite{Yukalov:2019nhu}
\bibitem{Yukalov:2019nhu}
V.~I.~Yukalov,
``Interplay between Approximation Theory and Renormalization Group,''
Phys. Part. Nucl. \textbf{50}, 141 (2019)
\href{https://doi.org/10.1134/S1063779619020047}{doi:10.1134/S1063779619020047}
\href{https://arxiv.org/pdf/2105.12176}{[arXiv:2105.12176 [hep-th]]}.
%24 citations counted in INSPIRE as of 28 Jun 2024


%\cite{Kneur:2007vj}
\bibitem{Kneur:2007vj}
J.~L.~Kneur, M.~B.~Pinto, R.~O.~Ramos and E.~Staudt,
``Updating the Phase Diagram of the Gross-Neveu Model in 2+1 Dimensions,''
Phys. Lett. B \textbf{657}, 136 (2007)
\href{https://doi.org/10.1016/j.physletb.2007.10.013}{doi:10.1016/j.physletb.2007.10.013}
\href{https://arxiv.org/pdf/0705.0673}{[arXiv:0705.0673 [hep-ph]]}.
%48 citations counted in INSPIRE as of 01 Jul 2024


%\cite{Kneur:2007vm}
\bibitem{Kneur:2007vm}
J.~L.~Kneur, M.~B.~Pinto, R.~O.~Ramos and E.~Staudt,
``Emergence of Tricritical Point and Liquid-Gas Phase in the Massless 2+1 Dimensional Gross-Neveu Model,''
Phys. Rev. D \textbf{76}, 045020 (2007)
\href{https://doi.org/10.1103/PhysRevD.76.045020}{doi:10.1103/PhysRevD.76.045020}
\href{https://arxiv.org/pdf/0705.0676}{[arXiv:0705.0676 [hep-th]]}.
%68 citations counted in INSPIRE as of 01 Jul 2024


%\cite{Kneur:2010yv}
\bibitem{Kneur:2010yv}
J.~L.~Kneur, M.~B.~Pinto and R.~O.~Ramos,
``Thermodynamics and Phase Structure of the Two-Flavor Nambu--Jona-Lasinio Model Beyond Large-$N_c$,''
Phys. Rev. C \textbf{81}, 065205 (2010)
\href{https://doi.org/10.1103/PhysRevC.81.065205}{doi:10.1103/PhysRevC.81.065205}
\href{https://arxiv.org/pdf/1004.3815}{[arXiv:1004.3815 [hep-ph]]}.
%49 citations counted in INSPIRE as of 01 Jul 2024


%\cite{Kneur:2013cva}
\bibitem{Kneur:2013cva}
J.~L.~Kneur, M.~B.~Pinto and R.~O.~Ramos,
``Phase Diagram of the Magnetized Planar Gross-Neveu Model Beyond the Large-$N$ Approximation,''
Phys. Rev. D \textbf{88}, 045005 (2013)
\href{https://doi.org/10.1103/PhysRevD.88.045005}{doi:10.1103/PhysRevD.88.045005}
\href{https://arxiv.org/pdf/1306.2933}{[arXiv:1306.2933 [hep-ph]]}.
%25 citations counted in INSPIRE as of 01 May 2024


%\cite{Pinto:1999py}
\bibitem{Pinto:1999py}
M.~B.~Pinto and R.~O.~Ramos,
``High Temperature Resummation in the Linear Delta Expansion,''
Phys. Rev. D \textbf{60}, 105005 (1999)
\href{https://doi.org/10.1103/PhysRevD.60.105005}{doi:10.1103/PhysRevD.60.105005}
\href{https://arxiv.org/pdf/hep-ph/9903353}{[arXiv:hep-ph/9903353 [hep-ph]]}.
%49 citations counted in INSPIRE as of 01 May 2024


%\cite{Pinto:1999pg}
\bibitem{Pinto:1999pg}
M.~B.~Pinto and R.~O.~Ramos,
``A Nonperturbative Study of Inverse Symmetry Breaking at High Temperatures,''
Phys. Rev. D \textbf{61}, 125016 (2000)
\href{https://doi.org/10.1103/PhysRevD.61.125016}{doi:10.1103/PhysRevD.61.125016}
\href{https://arxiv.org/pdf/hep-ph/9912273}{[arXiv:hep-ph/9912273 [hep-ph]]}.
%59 citations counted in INSPIRE as of 01 May 2024


%\cite{Farias:2008fs}
\bibitem{Farias:2008fs}
R.~L.~S.~Farias, G.~Krein and R.~O.~Ramos,
``Applicability of the Linear Delta Expansion for the $\lambda \phi^4$ Field Theory at Finite Temperature in the Symmetric and Broken Phases,''
Phys. Rev. D \textbf{78}, 065046 (2008)
\href{https://doi.org/10.1103/PhysRevD.78.065046}{doi:10.1103/PhysRevD.78.065046}
\href{https://arxiv.org/pdf/0809.1449}{[arXiv:0809.1449 [hep-ph]]}.
%32 citations counted in INSPIRE as of 01 May 2024


%\cite{Rosa:2016czs}
\bibitem{Rosa:2016czs}
D.~S.~Rosa, R.~L.~S.~Farias and R.~O.~Ramos,
``Reliability of the Optimized Perturbation Theory in the 0-Dimensional $O(N)$ Scalar Field Model,''
Physica A \textbf{464}, 11 (2016)
\href{https://doi.org/10.1016/j.physa.2016.07.067}{doi:10.1016/j.physa.2016.07.067}
\href{https://arxiv.org/pdf/1604.00537}{[arXiv:1604.00537 [hep-ph]]}.
%11 citations counted in INSPIRE as of 01 May 2024


%\cite{Farias:2021ult}
\bibitem{Farias:2021ult}
R.~L.~S.~Farias, R.~O.~Ramos and D.~S.~Rosa,
``Symmetry Breaking Patterns for Two Coupled Complex Scalar Fields at Finite Temperature and in an External Magnetic Field,''
Phys. Rev. D \textbf{104}, 096011 (2021)
\href{https://doi.org/10.1103/PhysRevD.104.096011}{doi:10.1103/PhysRevD.104.096011}
\href{https://arxiv.org/pdf/2109.03671}{[arXiv:2109.03671 [hep-ph]]}.
%4 citations counted in INSPIRE as of 01 May 2024


%\cite{Stevenson:1981vj}
\bibitem{Stevenson:1981vj}
P.~M.~Stevenson,
``Optimized Perturbation Theory,''
Phys. Rev. D \textbf{23}, 2916 (1981)
\href{https://doi.org/10.1103/PhysRevD.23.2916}{doi:10.1103/PhysRevD.23.2916}.
%1276 citations counted in INSPIRE as of 28 Jun 2024


%\cite{Kneur:2006ht}
\bibitem{Kneur:2006ht}
J.~L.~Kneur, M.~B.~Pinto and R.~O.~Ramos,
``Critical and Tricritical Points for the Massless 2D Gross-Neveu Model Beyond Large $N$,''
Phys. Rev. D \textbf{74}, 125020 (2006)
\href{https://doi.org/10.1103/PhysRevD.74.125020}{doi:10.1103/PhysRevD.74.125020}
\href{https://arxiv.org/pdf/hep-th/0610201}{[arXiv:hep-th/0610201 [hep-th]]}.
%39 citations counted in INSPIRE as of 01 Jul 2024


%\cite{Gandhi:1990yj}
\bibitem{Gandhi:1990yj}
S.~K.~Gandhi, H.~F.~Jones and M.~B.~Pinto,
``The Delta Expansion in the Large $N$ Limit,''
Nucl. Phys. B \textbf{359}, 429 (1991)
\href{https://doi.org/10.1016/0550-3213(91)90067-8}{doi:10.1016/0550-3213(91)90067-8}.
%68 citations counted in INSPIRE as of 29 Feb 2024


%\cite{Hyun:1994fb}
\bibitem{Hyun:1994fb}
S.~Hyun, G.~H.~Lee and J.~H.~Yee,
``Gaussian Approximation of the (2+1)-Dimensional Thirring Model in the Functional Schrodinger Picture,''
Phys. Rev. D \textbf{50}, 6542 (1994)
\href{https://doi.org/10.1103/PhysRevD.50.6542}{doi:10.1103/PhysRevD.50.6542}
\href{https://arxiv.org/pdf/hep-th/9406070}{[arXiv:hep-th/9406070 [hep-th]]}.
%12 citations counted in INSPIRE as of 29 Feb 2024


%\cite{Rosenstein:1990nm}
\bibitem{Rosenstein:1990nm}
B.~Rosenstein, B.~Warr and S.~H.~Park,
``Dynamical Symmetry Breaking in Four Fermi Interaction Models,''
Phys. Rept. \textbf{205}, 59 (1991)
\href{https://doi.org/10.1016/0370-1573(91)90129-A}{doi:10.1016/0370-1573(91)90129-A}.
%325 citations counted in INSPIRE as of 04 Jun 2024


%\cite{Kneur:2012qp}
\bibitem{Kneur:2012qp}
J.~L.~Kneur, M.~B.~Pinto, R.~O.~Ramos and E.~Staudt,
``Vector-Like Contributions from Optimized Perturbation in the Abelian Nambu--Jona-Lasinio Model for Cold and Dense Quark Matter,''
Int. J. Mod. Phys. E \textbf{21}, 1250017 (2012)
\href{https://doi.org/10.1142/S0218301312500176}{doi:10.1142/S0218301312500176}
\href{https://arxiv.org/pdf/1201.2860}{[arXiv:1201.2860 [nucl-th]]}.
%9 citations counted in INSPIRE as of 29 Feb 2024


%\cite{Restrepo:2014fna}
\bibitem{Restrepo:2014fna}
T.~E.~Restrepo, J.~C.~Macias, M.~B.~Pinto and G.~N.~Ferrari,
``Dynamical Generation of a Repulsive Vector Contribution to the Quark Pressure,''
Phys. Rev. D \textbf{91}, 065017 (2015)
\href{https://doi.org/10.1103/PhysRevD.91.065017}{doi:10.1103/PhysRevD.91.065017}
\href{https://arxiv.org/pdf/1412.3074}{[arXiv:1412.3074 [hep-ph]]}.
%28 citations counted in INSPIRE as of 01 May 2024

%\cite{Moshe:2003xn}
\bibitem{Moshe:2003xn}
M.~Moshe and J.~Zinn-Justin,
``Quantum field theory in the large N limit: A Review,''
Phys. Rept. \textbf{385}, 69-228 (2003)
doi:10.1016/S0370-1573(03)00263-1
[arXiv:hep-th/0306133 [hep-th]].
%429 citations counted in INSPIRE as of 14 Aug 2024

\end{thebibliography}
\end{document}